\RequirePackage{fix-cm}
\documentclass[smallextended]{svjour3}       
\smartqed  
\usepackage{graphicx}
\usepackage{hyperref}
\usepackage{natbib}
\usepackage{xcolor}
\usepackage{amssymb}
\usepackage{aas_macros}

\definecolor{patrick}{rgb}{0.75, 0.25, 0.0}

\definecolor{ineke}{rgb}{0.7, 0.0, 0.5}

\title{Magnetohydrodynamic waves in open coronal structures}

\titlerunning{MHD Waves in open structures}        

\author{D. Banerjee \and
        S. Krishna Prasad \and V. Pant \and J.A. McLaughlin \and P. Antolin \and N. Magyar \and L. Ofman \and H. Tian \and T. Van Doorsselaere \and I. De Moortel \and T. J. Wang
}

\authorrunning{D. Banerjee et al.} 

\institute{D. Banerjee \at
             Aryabhatta Research Institute of Observational Sciences, Nainital-263002, India\\
             Indian Institute of Astrophysics, Koramangala, Bangalore 560034, India\\
             Center of Excellence in Space Sciences India, Indian Institute of Science Education and Research Kolkata, Mohanpur 741246, West Bengal, India\\
              \email{dipu@aries.res.in}           
           \and 
		  S. Krishna Prasad \at
		    Centre for mathematical Plasma Astrophysics, Department of Mathematics, KU Leuven, Celestijnenlaan 200B, B-3001 Leuven, Belgium
		   \and	
           V. Pant \at
           Aryabhatta Research Institute of Observational Sciences, Nainital-263002, India
           \and
          James A. McLaughlin [0000-0002-7863-624X] \at
            Northumbria University, Newcastle upon Tyne, NE1 8ST, UK
           \and
          P. Antolin \at
            Department of Mathematics, Physics and Electrical Engineering, Northumbria University, Newcastle Upon Tyne NE1 8ST, UK
           \and
          N. Magyar \at
            Centre for Fusion, Space and Astrophysics, Physics Department, University of Warwick, Coventry CV4 7AL, UK
           \and
          L. Ofman \at
            The Catholic University of America and NASA Goddard Space Flight Center, Code 671, Greenbelt, MD 20771, USA
            \and    
          H. Tian \at
            School of Earth and Space Sciences, Peking University, Beijing 100871, China\\
            Key Laboratory of Solar Activity, National Astronomical Observatories, Chinese Academy of Sciences, Beijing 100012, China
            \and
          T. Van Doorsselaere \at
            Centre for mathematical Plasma Astrophysics, Department of Mathematics, KU Leuven, Celestijnenlaan 200B, B-3001 Leuven, Belgium
           \and
          Ineke De Moortel \at 
            School of Mathematics and Statistics, University of St. Andrews, St. Andrews, KY16 9SS, U.K. \\ Rosseland Centre for Solar Physics, University of Oslo, PO Box 1029  Blindern, NO-0315 Oslo, Norway
           \and
          T. J. Wang \at
            The Catholic University of America and NASA Goddard Space Flight Center, Code 671, Greenbelt, MD 20771, USA}

\date{Received: date / Accepted: date}

\begin{document}

\maketitle

\begin{abstract}
Modern observatories have revealed the ubiquitous presence of magnetohydrodynamic waves in the solar corona. The propagating waves (in contrast to the standing waves) are usually originated in the lower solar atmosphere which makes them particularly relevant to coronal heating. Furthermore, open coronal structures are believed to be the source regions of solar wind, therefore, the detection of MHD waves in these structures is also pertinent to the acceleration of solar wind. Besides, the advanced capabilities of the current generation telescopes have allowed us to extract important coronal properties through MHD seismology. The recent progress made in the detection, origin, and damping of both propagating slow mangetoacoustic waves and kink (Alfv\'{e}nic) waves is presented in this review article especially in the context of open coronal structures. Where appropriate, we give an overview on associated theoretical modelling studies. A few of the important seismological applications of these waves are discussed. The possible role of Aflv\'{e}nic waves in the acceleration of solar wind is also touched upon.

\end{abstract}

\section{Introduction}
\label{intro}
Dense and hot plasma structures are a common sight in the solar corona. These structures are mostly linear (in the sense that their lengths are much longer than their transverse scales) and are believed to trace local magnetic field lines. Some of them are closed, meaning both ends of the structure anchored on the Sun (footpoints) are clearly visible while others that are extended to very large distances concealing their connection from one foot point to other, are considered open. Ray-like structures visible in the polar regions of the Sun, called polar plumes, are open structures. The structures that form part of active regions are typically closed and are called coronal loops. It is also common to have ``open loops'' or ``extended loops'', referring to the loops that are part of active regions but their connectivity to the other foot point is not clear. Extended loops originating in a quiet-Sun network region (or a plage region) are called network plumes or on-disk plume-like structures. Because of the constant churning experienced by all these structures due to convective motions on the surface, it is natural to expect creation of different magnetohydrodynamic (MHD) wave modes which are believed to play a significant role in maintaining the million degree temperatures in the corona. We refer the reader to a review by \citet{2020SSRv..216..140V} in this issue, to find our current understanding on the contribution of MHD waves to coronal heating.

The launch of high-resolution telescopes such as SoHO and TRACE in the late 90s has led to the discovery of different MHD wave modes in the solar corona. Both longitudinal slow magnetoacoustic waves and transverse kink/Alfv\'enic waves were found to exist. Modern telescopes such as Hinode, SDO, and IRIS alongside several ground-based telescopes (for e.g., SST, DST, GREGOR, GST, DKIST) offer us even better resolution in addition to providing increased spatial, temporal, and atmospheric coverage. These extraordinary capabilities have not only expanded the detection of different MHD waves but also improved our knowledge on their characteristic properties. Indeed, as many researchers have shown, current observations are advanced enough to compare the observed wave properties with theoretical models and extract several key parameters through coronal seismology - a newly emergent field in solar physics, thanks to these advances \citep[e.g.,][]{1999Sci...285..862N, 2001A&A...372L..53N, 2012RSPTA.370.3193D, 2020ARA&A..58..441N}. 

Slow magnetoacoustic waves (or simply slow waves) have been detected in different coronal structures including polar plumes \citep[e.g.,][]{1997ApJ...491L.111O, 1998ApJ...501L.217D}, active region fan loops \citep[e.g.,][]{2002A&A...387L..13D, 2006ApJ...643..540M} and other network or plage related plume-like structures \citep[e.g.,][]{1999SoPh..186..207B, 2000A&A...355L..23D}. Standing slow waves were also found in hot flare loops \citep[e.g,][]{2011SSRv..158..397W}. A dedicated review on standing waves has been presented by \citet{2021SSRv..217...34W} in this issue. Recent studies have shown the origin of propagating slow waves in the lower solar atmosphere \citep[e.g.,][]{2009A&A...505..791S, 2011ApJ...728...84B, 2012ApJ...757..160J, 2015ApJ...812L..15K, 2016ApJ...830L..17Z}, revealed their damping to be frequency-dependent albeit with some inconsistencies with theory \citep[e.g.,][]{2014ApJ...789..118K}, and also highlighted that thermal conduction might not be the dominant cause for their damping as previously thought \citep[e.g.,][]{2015ApJ...811L..13W}. The high-resolution observations also uncovered a few problems, one of which is the possible ambiguity in distinguishing the slow waves from high-speed quasi-periodic flows. Especially, the spectroscopic observations revealed a more complex picture, with perturbations not only in the line intensities but also in other parameters such as the Doppler velocities, line widths and red-blue asymmetry leading to the debate on interpretation in terms of waves \citep[e.g.,][]{2009A&A...503L..25W, 2009A&A...499L..29B, 2010ApJ...721..744K, 2010ApJ...713..573M, 2011A&A...528L...4K, 2011ApJ...734...81M, 2014ApJ...793..117S, 2015RAA....15.1027D, 2016RAA....16...93J, 2016ApJS..224...30Y, 2018ApJ...853..134M, 2021ApJ...909..202C} or flows \citep[e.g.,][]{2007Sci...318.1585S, 2007ApJ...667L.109D, 2008ApJ...678L..67H, 2010ApJ...722.1013D, 2010A&A...516A..14H, 2010RAA....10.1307G, 2010ApJ...715.1012B, 2011ApJ...730...37U, 2011ApJ...736..130T, 2012ApJ...759..144T, 2012ApJ...760...82S, 2016ApJ...829L..18B, 2021SoPh..296...47T}. Despite this, a number of seismological applications of slow waves have been proposed including the determination of plasma temperature, loop geometry, magnetic field, thermal conduction and polytropic index \citep{2009ApJ...697.1674M, 2009A&A...503L..25W, 2011ApJ...727L..32V, 2014A&A...561A..19Y, 2015ApJ...811L..13W, 2016NatPh..12..179J, 2017ApJ...834..103K, 2018SoPh..293....2D, 2018ApJ...868..149K, 2018ApJ...860..107W, 2019ApJ...886....2W}. More recent models discuss the inclusion of unknown coronal heating function in the  MHD equations in a parameterised form as a function of the equilibrium parameters such as density and temperature in order to constrain it using the observations of slow waves \citep{2019A&A...628A.133K, 2019PhPl...26h2113Z, 2020A&A...644A..33K, 2021SoPh..296...20P}.

Different MHD wave modes are ubiquitously observed in the solar atmosphere by space and ground-based instruments \citep{2007SoPh..246....3B}.
It is worth noting that early studies since the era of {\it Skylab} have interpreted the observed nonthermal broadening of spectral lines to be the signature of Alfv\'en waves in polar coronal holes \citep[][and references therein]{1990ApJ...348L..77H,1998A&A...339..208B,2009A&A...501L..15B}. However, the observational resolution in that era was not good enough to disentangle different wave modes (torsional Alfv\'en wave or homogenous Alfv\'en wave) and the structure of the solar plasma. After the launch of high resolution instruments, there is an ample evidence that the solar plasma is inhomogeneous. Hence, the term Alfvénic (also called kink) better acknowledges the rich and complex nature of the coupled wave system within a realistic, continuous, inhomogeneous magnetic medium. 

Direct signatures of kink waves were first noted in coronal loops \citep{1999Sci...285..862N,1999ApJ...520..880A}. Both decaying and decayless oscillations have been observed in different solar structures \citep{2013A&A...552A..57N,2015A&A...583A.136A}. It has been reported that most of the kink oscillations are excited when coronal loops are displaced from their mean position by the lower coronal eruptions \citep{2015A&A...577A...4Z} including coronal jets \citep{2016SoPh..291.3269S}. After the launch of CoMP and SDO, propagating kink waves were detected in the polar regions of the Sun \citep[][and references therein]{2007Sci...317.1192T,2009ApJ...697.1384T,2014ApJ...790L...2T}.
It has been believed that the sources of these propagating kink waves in polar coronal holes are the supergranular motions. Recently, \citep{2016ApJ...828...89M, 2019NatAs...3..223M} has shown that p-modes could be responsible for exciting observed kink oscillations in the solar atmosphere. While \citet{2016A&A...591L...5N} showed that the observed dependence of the displacement and velocity amplitudes on the oscillation periods does not have any peak at a certain frequency. This casts doubts over the interpretations that p-modes are responsible for observed kink waves in the solar corona. Also, kink waves (standing and propagating) can be used to infer cross density structures and magnetic field using seismology inversion techniques \citep[][and references therein]{2011ApJ...736...10S,2019ApJ...876..106T,2020ApJ...894...79W,2020Sci...369..694Y}. Though, a unique solution of the inversion does not exist \citep[see][and references therein]{2019A&A...622A..44A} except when the density profile of the inhomogeneous layer is kept fixed \citep{2016A&A...589A.136P}.
Transverse waves are believed to carry energy flux enough to heat the quiet Sun\citep{2011Natur.475..477M}. While \citet{2014ApJ...790L...2T} claimed that transverse waves carry insufficient energy to accelerate solar winds in polar coronal holes, \citet{2014Sci...346A.315T} reported that transverse waves observed in the transition regions have enough energies to accelerate the solar wind. Thus, more studies using upcoming space and ground-based instruments are needed to understand the role of transverse waves in accelerating solar winds in the polar coronal holes. 

In this review article, we give an overview of some of these recent advancements in the observations of both slow waves and Alf\'{e}nic waves with an emphasis on open coronal structures. We also present relevant theoretical/numerical modelling studies wherever appropriate. We discuss some of the latest advances in the study of slow waves in Section \ref{slow_waves}, and that in Alfv\'{e}nic or kink waves in Section \ref{alfven_waves}, followed by a description of the role of Alfv\'{e}nic waves in the solar wind acceleration in Section \ref{solar_wind} and finally provide a brief summary of all the results along with some future directions in Section \ref{summary}.

\section{Slow magnetoacoustic waves} 
\label{slow_waves}
Observations of propagating slow magnetoacoustic waves in the solar atmosphere have generated increased attention over the past decades. The continuous multi-wavelength capture of the full solar disk at high spatial and temporal resolutions by SDO/AIA, has enabled us to perform detailed studies on the slow wave propagation and dissipation characteristics. The perturbations due to slow waves appear as slanted ridges of alternating brightness (or Doppler velocity) in time-distance maps that are commonly constructed to study the evolution of a specific coronal structure. These perturbations are generally referred to as `propagating disturbances (PDs)' or `propagating coronal disturbances (PCDs)'. There has been a debate in the recent years proclaiming the observed PDs could also be generated by high-speed quasi-periodic upflows \citep[e.g.,][and references cited therein]{2010ApJ...722.1013D, 2010A&A...510L...2M, 2011ApJ...727L..37T, 2011ApJ...738...18T, 2012ApJ...749...60M} while their interpretation in terms of slow waves \citep[e.g.,][and references cited therein]{2009A&A...503L..25W, 2010ApJ...724L.194V, 2011A&A...528L...4K,  2012ASPC..456...91W, 2012A&A...546A..93G} is not unique. Importantly, some spectroscopic observations of PDs from \textit{Hinode}/EIS have revealed periodic blueward asymmetry in the line profiles suggesting the presence of periodic upflows \citep{2010ApJ...722.1013D}. However, \citet{2010ApJ...724L.194V} have later demonstrated that such asymmetry in the line profiles is an intrinsic feature of upward propagating slow waves.  Although the issue has not been completely resolved yet \citep[e.g.,][and references cited therein]{2012RSPTA.370.3193D, 2015SoPh..290..399D, 2017ApJ...845L..18D}, it has been largely agreed that both periodic waves and aperiodic flows coexist near the foot points but the wave signatures dominate at larger distances along the structure \citep[e.g.,][and references cited therein]{2011ApJ...737L..43N, 2012ApJ...759..144T,2012ApJ...754..111O,2013ApJ...775L..23W, 2015ApJ...807...71P, 2015ApJ...815L..16S}. The discussion in this article presumes this behaviour and considers PDs synonymous with slow waves unless mentioned otherwise explicitly. Interested readers may refer to \cite{2012RSPTA.370.3193D}, \cite{2016GMS...216..419B}, and \cite{2016GMS...216..395W}, for further details on this debate. One must also note that the extended loop structures associated with active regions are considered as open structures in this article in the sense of the wave propagation but not magnetic connectivities that are unclear in observations. This is a valid assumption since the extent of propagation of slow waves is in general much shorter compared to the length (visible extent) of the associated structures. In the following, we discuss the progress made in the last few years in understanding the origin of the slow waves, their damping behaviour in the solar corona, and finally illustrate some of their seismological applications.

\subsection{Source of oscillations}
Slow waves are regularly observed in a variety of coronal structures. Both standing and propagating versions of these waves have been found although the standing waves are thought to be locally generated \citep{2011SSRv..158..397W, 2013ApJ...779L...7K}, whereas the propagating waves originate in the lower solar atmosphere \citep{2009A&A...505..791S, 2011ApJ...728...84B, 2012ApJ...757..160J, 2015ApJ...812L..15K}. Moreover, the propagating waves have been shown to be ubiquitous in the solar corona \citep{2012A&A...546A..50K, 2018ApJ...853..145M}. A wide range of periodicities are observed, but in general, longer periods ($\sim$ tens of minutes) are dominant in the polar regions \citep{2009A&A...499L..29B, 2011A&A...528L...4K} while the active regions are replete with shorter periods ($\sim$ minutes) \citep{2012SoPh..281...67K, 2018ApJ...856L..16W}. However, the longer periods ($>$ 10 min) are difficult to penetrate into the solar corona unless the magnetic field is highly inclined \citep{2015ApJ...809L..17J}. Therefore, even though the general characteristic properties of these waves appear broadly similar across all regions, it is quite possible that they have a different origin in active regions and polar regions. Here, we discuss some recent results on possible source locations across these two regions. It may be noted that longer periods ($\sim$ 25 min) are also found in active regions \citep{2009A&A...503L..25W, 2011A&A...533A.116Y} and similarly, shorter periods ($\sim$ 5 min) are not totally non-existent in the polar regions \citep{2014ApJ...789..118K} but the distinction here is based on their dominant presence.

\subsubsection{Active regions}
Early observations using high-resolution imaging data from \textit{Transition Region and Coronal Explorer} \citep[TRACE;][]{1999SoPh..187..229H} have revealed that the slow wave perturbations in loops that are rooted in a sunspot umbra display 3-minute periods whereas those in loops anchored in plage-related regions display 5-minute periods \citep{2002A&A...387L..13D}. These periods are coincident with that of the oscillations found in the lower atmosphere at the corresponding locations. It was therefore suggested that the driver of the coronal slow waves lies in the lower atmosphere near the footpoints of the supporting structure. \cite{2006ApJ...643..540M} used spectroscopic data corresponding to the transition region of a sunspot along with the co-temporal coronal images of the region, to show the propagation of slow waves from the transition region to corona. The authors further conjectured that these oscillations are driven by the global oscillations of the Sun, i.e., the photospheric $p$-modes. It is possible that the $p$-modes act as a broadband driver and the local magnetic configuration aids the leakage of the natural frequencies in the vicinity of the acoustic cutoff in the form of upward propagating slow waves \citep{2011ApJ...728...84B}.

The amplitude of propagating slow waves is not constant but appears to be modulated over time. Assuming the local medium is relatively stable, the temporal changes in a propagating wave implies changes in the driver properties. Utilising this idea, \cite{2015ApJ...812L..15K} traced the coronal slow waves across different layers of the solar atmosphere to identify their driver. High-resolution multi-wavelength observations of a sunspot, acquired from both ground- and space-based telescopes in 9 different channels, are used to perform this study.
\begin{figure}
\includegraphics[width=0.25\textwidth, trim={1.6cm 0 0 0}, clip]{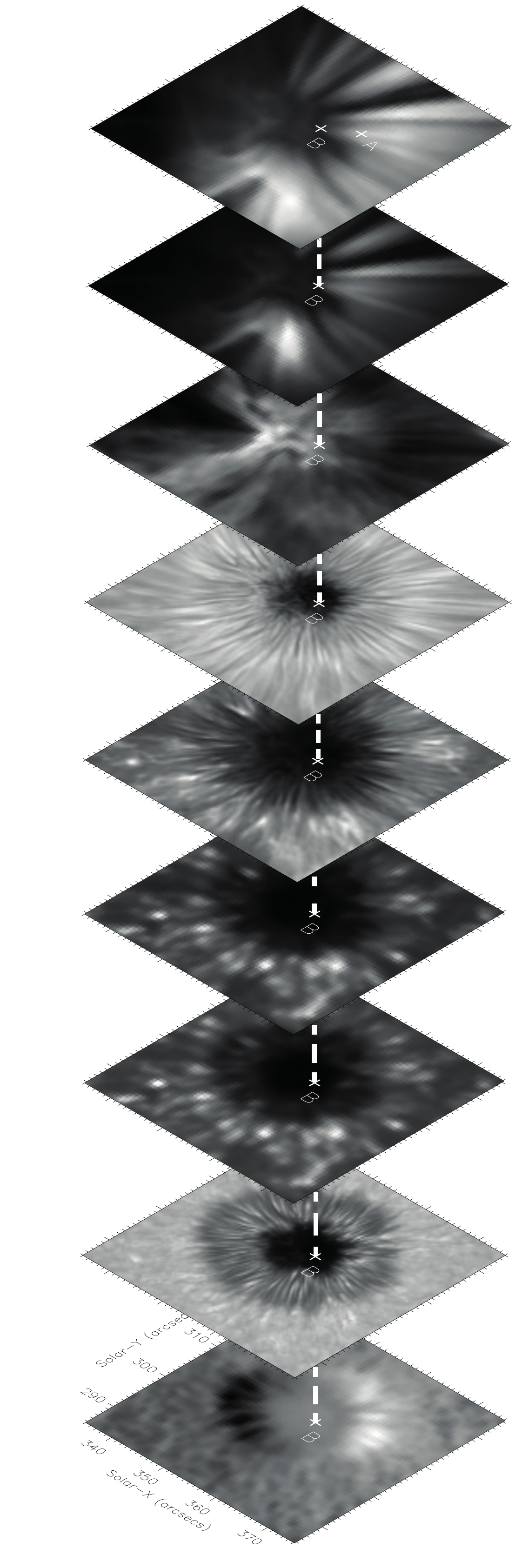}
\includegraphics[width=0.75\textwidth]{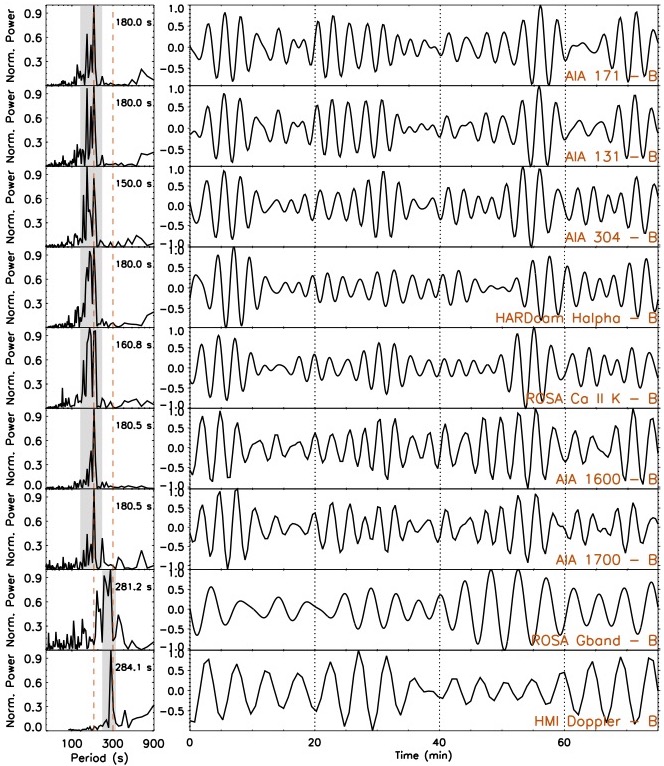}
\caption{Structure of the leading spot of NOAA 11366 across 9 different wavelength channels captured on 10 December 2011 by a suite of telescopes based on ground and space through coordinated observations. The vertical dashed line connects the footpoint location of a fan loop (marked as `B') rooted in the umbra across different channels. The central panels display the corresponding Fourier power spectra of light curves from location `B'. The peak periodicities are listed in the individual panels. The vertical dashed lines denote the locations of 3 and 5 minute periods. The filtered light curves allowing frequencies only within a narrow band (indicated in grey in the central panels) around the peak frequency are shown in the right panels. The names of the individual channels are also listed. Adapted from \cite{2015ApJ...812L..15K}}
\label{kpfig1}
\end{figure}
Fig.~\ref{kpfig1} (left panel) displays the structure of sunspot across 9 wavelength channels. The fan loop structure is evident in the top two panels characterising coronal emission. The location `B' corresponds to the foot point of a selected loop which displays propagating slow waves and the vertical dashed line connects this location across all the channels. The central panels display the corresponding Fourier power spectra of the light curves extracted from this location. The peak periods are listed in each panel. The vertical dashed lines in red denote the locations of the 3 and 5 minute periods. The shift in periodicity from 5 min in the photosphere (bottom two panels) to 3 min in the chromosphere/transition region is long-known with multiple explanations already provided in the literature \citep{1983SoPh...82..369Z, 1991A&A...250..235F}. The grey bands over the spectra delineate the widths of the band-pass filter applied on the original data to obtain the light curves displayed in the right panels. The identification labels for the individual channels are listed at the bottom of the light curves. As may be seen from the light curves, the amplitude of the oscillations is periodically varying with approximately similar behaviour across all the channels. The modulation period is found to be in the range of 20$-$27 min. These correlated amplitude variations across different layers imply that they are clearly non local and therefore should be associated with the driver. \cite{2015ApJ...812L..15K} have further demonstrated that similar modulation periods are found in the global acoustic oscillations at non-magnetic locations outside the sunspot and are fully consistent with those seen in coronal oscillations. The modulation itself is caused by the simultaneous presence of waves with closely spaced frequencies (as may be seen in the Fourier power spectra in Fig.~\ref{kpfig1}) as noted previously by several authors \citep[e.g.,][]{2006ApJ...643..540M} and is a characteristic feature of global oscillations. These results compelled the authors to conclude that the coronal slow waves are externally driven by the global solar oscillations ($p$-modes) in the photosphere. A recent study by \cite{2020A&A...638A...6S} also shows similar amplitude modulation in slow waves observed in a sunspot across different layers of the solar atmosphere possibly confirming the wider applicability of this interpretation. Utilising high-resolution multi-wavelength observations of a sunspot from 16 different channels, \citet{2016ApJ...830L..17Z} studied the propagation of slow waves from photosphere to corona. These authors employed a time-distance helioseismic technique to cross-correlate and track the waves across different layers of the solar atmosphere. They analysed the propagation patterns of different frequencies and compared them with a numerical model to demonstrate that the observed slow waves in the corona originate from a source few megameters beneath the photosphere. It was further suggested that the interaction of $p$-modes with the sub-surface magnetic fields of sunspots can possibly explain the underlying source.

By combining high-spatial and temporal resolution data from the ground-based Dunn Solar Telescope (DST) with the Extreme Ultraviolet (EUV) imaging data from the SDO/AIA, \cite{2012ApJ...757..160J} found $\approx$3 minute oscillations across different layers of the solar atmosphere. It has been further demonstrated that the coronal fan structures, where propagating slow waves are observed, are rooted at specific locations in the umbra identified by small-scale bright features called Umbral Dots (UDs) and the oscillation power at these locations in the 3-minute period band is at least three orders of magnitude higher than the surrounding umbra. The remarkable correspondence between the oscillation properties across different layers led the authors to suggest that the source of the coronal slow waves is located in the umbral photosphere. \cite{2017ApJ...836...18C} studied velocity oscillations in the photosphere of a sunspot umbra and observed similar enhancements in the 3 minute oscillation power in the vicinity of umbral dots and light bridges. Additionally, they found that these oscillations occur independent of the 5 minute oscillations seen in the rest of the umbra and display an upward propagating nature. Based on these results, in addition to the possibility that umbral dots and light bridges are locations of magnetoconvection, the authors speculated that 3 minute oscillations in sunspots are locally driven through turbulent convection as predicted previously by certain models \citep{1967SoPh....2..385S, 1993ApJ...404..372L}. More recently, \cite{2019ApJ...879...67C} identified different oscillation patterns in the umbral photosphere, all centred on umbral dots that exhibit large morphological and dynamical changes. These authors again concluded magnetoconvection as a possible source of sunspot oscillations. \cite{2014AstL...40..576Z} \& \cite{2020ApJ...888...84S} also find the sunspot oscillations are localised over individual cells of few arcseconds, however, to explain the multiple peaks in the oscillation spectra and their spatial distribution, the authors suggest the existence of a sub-photospheric resonator \citep{1984PAZh...10...51Z} that could drive these oscillations. 

\subsubsection{Polar regions}
As stated earlier, it is not trivial for long-period slow waves to propagate from photosphere to the outer solar atmosphere unless some special conditions are met. The ambiguity between wave and flow interpretations for PDs further complicates the issue. Using spectroscopic data from \textit{Hinode}, \cite{2011ApJ...737L..43N} found that line profiles near the base of a structure display flow-like signatures while those at larger distances are compatible with slow waves. The authors further concluded that both waves and flows may originate from unresolved explosive events in the lower corona. \cite{2012ApJ...754..111O} and \cite{2013ApJ...775L..23W} performed 3D MHD simulations demonstrating that impulsive flow pulses injected at the base of a coronal loop naturally excite slow waves which propagate along the loop while the flows themselves rapidly decelerate with height. \cite{2012A&A...546A..93G} studied PDs of 14.5 min periodicity observed in a south polar coronal hole using spectroscopic data. They found that the spectral parameters of Ne{\,}\textsc{viii} 770{\,}{\AA} line exhibit correlated periodic fluctuations in intensity and Doppler shift with no corresponding signatures in line width. Besides, the spectral line profiles did not show any visible asymmetry compelling the authors to interpret the observed PDs as slow waves. The co-temporal data from N{\,}\textsc{iv} 765{\,}{\AA} line, representing the transition region, displays periodic enhancements in line width near the base of the PDs with no associated variations in the other two line parameters. Based on these results, the authors suggest that the slow waves observed in the investigated polar region may be triggered by small-scale explosive events in the lower atmosphere. The base of a network plume, where PDs were observed, was examined in detail by \cite{2015ApJ...807...71P} using both imaging and spectroscopic data from SDO/AIA and IRIS. Quasi-periodic brightenings were found near the base with spectral profiles occasionally displaying Doppler shifts with large deviations from the mean coincident with enhanced line widths and intensity consistent with a flow-like behaviour. The PDs observed along the plume were, however, compatible with slow wave properties. The brightenings were interpreted as due to small scale jet-like features \citep[e.g.,][]{2014Sci...346A.315T,2014ApJ...787..118R} and their correlation with the PDs, according to the authors, indicate them as a possible driver for propagating slow waves observed along the plume.

A four-hour-long image sequence, capturing the evolution of a north polar coronal hole at high spatial and temporal resolutions by SDO/AIA, is analysed by \cite{2015ApJ...809L..17J} to investigate the relation between the PDs observed in the corona and the spicules seen in the lower atmosphere. A direct correspondence between these two phenomena is detected in the composite images constructed by overlaying the lower-atmospheric 304{\,}{\AA} channel data on top of the coronal 171{\,}{\AA} channel data. 
\begin{figure}
\includegraphics[width=0.5\textwidth]{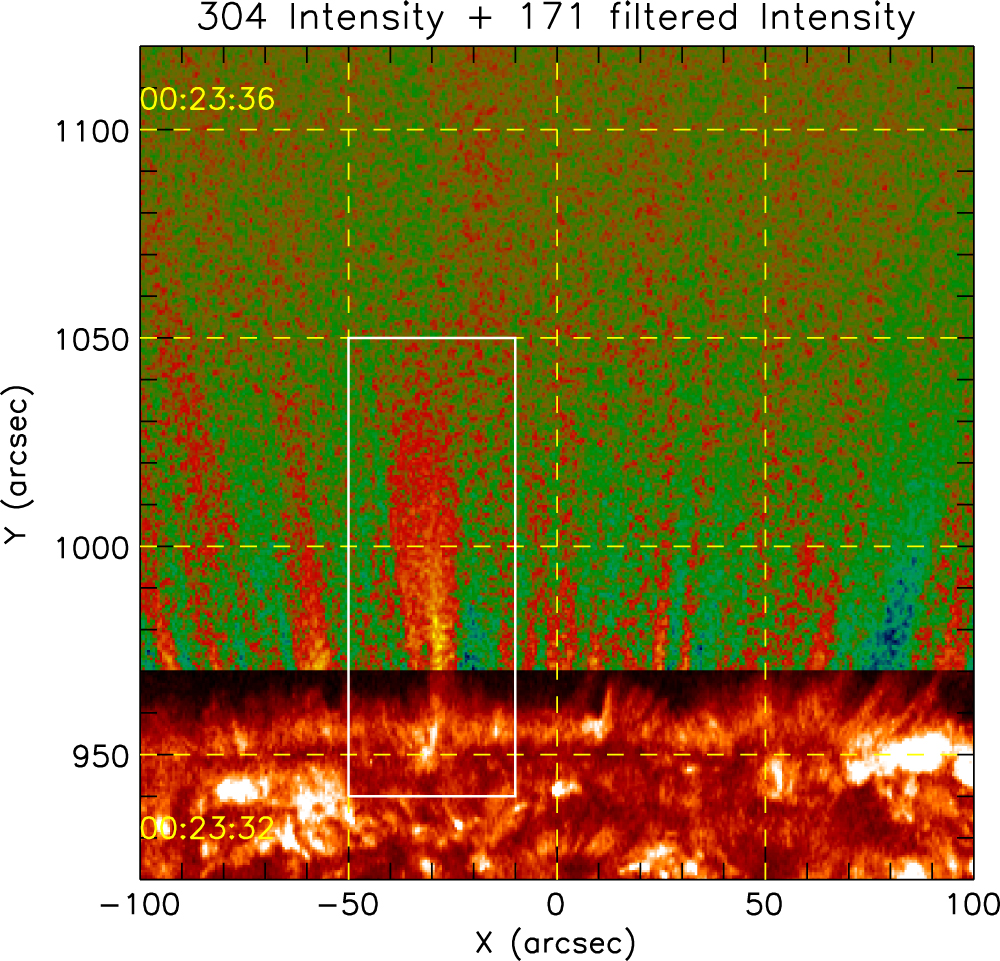}
\includegraphics[width=0.47\textwidth]{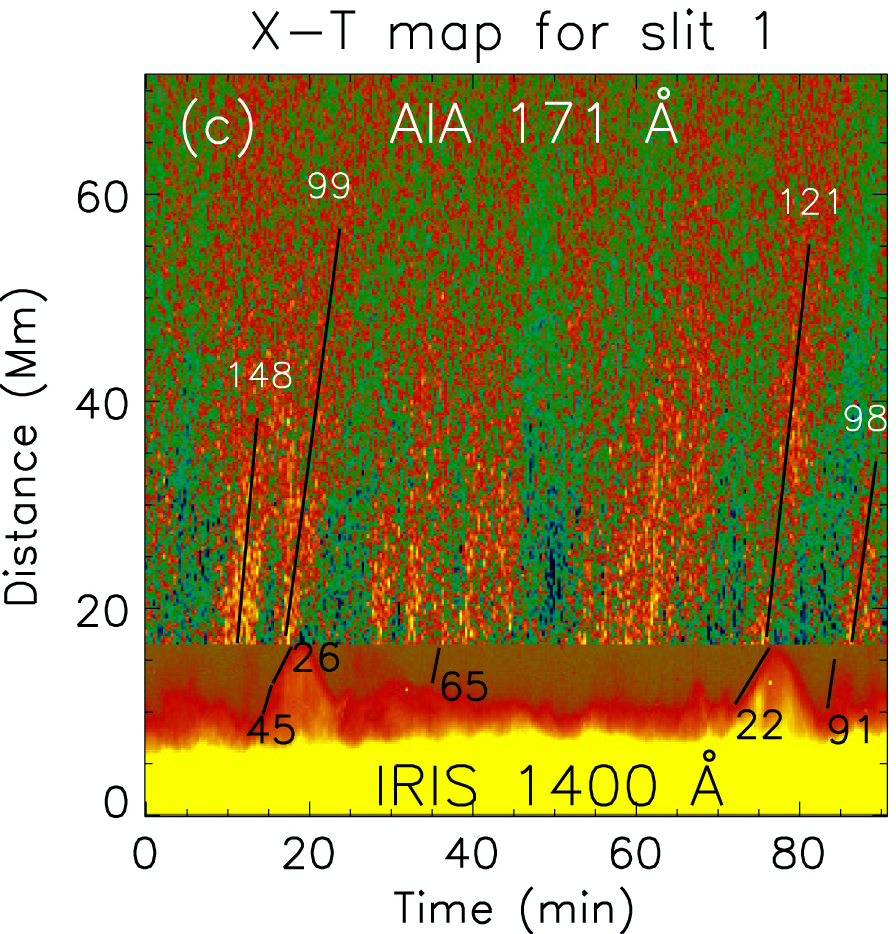}
\caption{\textit{Left}: A section of a north polar coronal hole of the Sun captured by SDO/AIA on 6 August, 2010. The image is a composite with the bottom and top parts displaying the emission observed in 304{\,}{\AA} and 171{\,}{\AA} channels, respectively. From \cite{2015ApJ...809L..17J}. \textit{Right}: A composite time-distance map with the bottom part displaying the rise and fall of spicules as seen in IRIS 1400{\,}{\AA} channel and the top part displaying the PDs as seen in AIA 171{\,}{\AA} channel. The slanted black lines tracking the individual features give an estimate of the propagation speed through their slope. The measured speeds (in km{\,}s$^{-1}$) are listed next to these lines. From \cite{2015ApJ...815L..16S}.}
\label{kpfig2}
\end{figure}
Fig.~\ref{kpfig2} shows one such image from their work in the left panel. While noting that a one-to-one correspondence is not always possible because of the overlapping structures along the line of sight especially near the limb, the authors demonstrate that the periodic behaviour of PDs is because of the repeated occurrence of spicules at the same location. Based on the observed speeds, it is further claimed that both type I and type II spicules contribute equally to the generation of PDs. The exact nature of the PDs is not explicitly addressed but the authors appeared to consider them as mass motions. \cite{2015ApJ...815L..16S} performed a similar study by combining the high-resolution slit-jaw images of a south polar region from IRIS with the co-temporal multi-wavelength data from SDO/AIA. Data from IRIS 2796{\,}{\AA}, and 1400{\,}{\AA} channels along with the AIA 304{\,}{\AA}, 171{\,}{\AA}, and 193{\,}{\AA} channels were used in this study. All these channels, when combined, encompass the observations of solar atmosphere from chromosphere, through transition region, to corona. A correspondence between the spicules in the lower atmosphere and the PDs in the corona is again observed. Furthermore, it is found that the beginning of each PD coincides with the rise phase of a spicule. Brightenings are observed in coronal channels, co-temporal to the fall of spicular material, following which another PD is initiated. A composite time-distance map describing the rise and fall of spicules (as seen in IRIS 1400{\,}{\AA} channel) and the simultaneous propagation of PDs (as seen in AIA 171{\,}{\AA} channel) is shown in the right panel of Fig.~\ref{kpfig2}. Additionally, the authors measured propagation speeds of PDs in 171{\,}{\AA} and 193{\,}{\AA} channels but could not determine if there is any temperature-dependent propagation \citep{2012SoPh..279..427K} because of the large uncertainties in their estimation. However, the presence of alternate bright and dark ridges and their continued existence at large altitudes led the authors to favour a slow-wave interpretation for the observed PDs. Finally, \cite{2015ApJ...815L..16S} conjecture that both spicules and slow waves are simultaneously driven by a reconnection-like process in the lower atmosphere, of which the cold spicular material falls back while the waves continue to propagate upward. Later, based on 1D MHD modelling, \citet{2016ApJS..224...30Y} show that finite-lifetime stochastic transients launched at the base of a magnetic structure can generate quasi-periodic oscillations with signatures similar to PDs along the structure. Gaussian pulses with the typical spicular timescales are used to simulate the transients in their study. An empirical atmospheric model is considered and depending on the smoothness of the thermal structure (with or without chromospheric resonant cavity), it was shown that both long and short-period PDs could be reproduced corresponding to those observed in polar plumes and active region loops, respectively. A more recent study by \citet{2019Sci...366..890S} has further illustrated that the connection between spicules and PD-like coronal moving features is very common even in on-disk observations. \\

\subsection{Damping/Dissipation properties}
The slow waves are observed to fall very rapidly in amplitude with their travelling distance along the solar coronal structures. Various physical mechanisms including thermal conduction, compressive viscosity, and optically thin radiation, which could affect the wave amplitude, were identified and studied through theoretical modelling in the past (see a complete review by \citet{2021SSRv..217...34W}, in this issue). The results indicated, for typical coronal conditions, that thermal conduction is a strong contributor to the observed damping of slow waves  and the effects of viscosity and radiation are negligible \citep{2002ApJ...580L..85O,2003A&A...408..755D, 2004A&A...415..705D,2018AdSpR..61..645P}. Later studies revealed exceptions to this scenario in some special conditions, e.g., the dissipation by thermal conduction is insufficient to explain the rapid damping of oscillations with longer periods in typical coronal loops \citep{2011ApJ...734...81M}, while compressive viscosity may become important when the loop is super-hot and with very low density \citep[see][]{2006SoPh..236..127P, 2007SoPh..246..187S, 2021SoPh..296...20P}, or when anomalous transport occurs in hot flaring plasma \citep{ 2019ApJ...886....2W}.

By constructing powermaps in different period ranges, \cite{2012A&A...546A..50K} studied the distribution of oscillatory power in three different open structures, namely, the active region fan loops, network plumes and the polar plumes. It is found that the oscillatory power at longer periods is significant up to larger distances in all the structures indicating a frequency-dependent damping. Furthermore, the measured amplitudes and damping lengths are shorter in hotter wavelength channels. These results are shown to be compatible with a propagating slow wave model incorporating thermal conduction as the main damping mechanism. The frequency dependence has been further explored by \cite{2014ApJ...789..118K}, where the authors attempted to find the exact quantitative dependence of damping length on the oscillation period. For each identified period, the Fourier amplitudes were computed as a function of distance, by fitting with an exponential decay function, from which the associated damping length is derived. Exploiting the simultaneous existence of multiple oscillation periods and combining the results from similar structures, the quantitative dependence of damping length on oscillation period is examined. Considering a power-law dependence of $L_d \propto P^{\alpha}$, where $L_d$ is the damping length and $P$ is the oscillation period, the $\alpha$ values (power-law indices) were obtained as 0.7$\pm$0.2 for the active region loops and network plumes and $-$0.3$\pm$0.1 for the polar region structures using AIA 171{\,}{\AA} channel. The corresponding values for the AIA 193{\,}{\AA} channel are 1.7$\pm$0.5, and $-$0.4$\pm$0.1, respectively. Assuming a 1D linear wave model, the authors also derived the expected theoretical dependences for different damping mechanisms, according to which, the $\alpha$ value should lie between 0 and 2. Notably, for the case of weak thermal conduction, the expected $\alpha$ value is 2 implying a steeper dependence and for the case of strong thermal conduction, this value should be 0 indicating no dependence or uniform damping across all periods. In reality, an intermediate value is possible depending on the strength of the thermal conduction but the obtained values are clearly deviant, especially those from polar regions, uncovering a large discrepancy between the observations and the theory of slow waves. It may be noted that the theoretical dependences ($\alpha$ values) are calculated under the assumption that only a single dissipation mechanism works (or perhaps dominant). Indeed, the negative slopes obtained for polar regions would mean stronger damping at longer periods which is not expected from any of the dissipation mechanisms studied by \cite{2014ApJ...789..118K} in the linear regime. Also, these results appear to be in direct contradiction with \cite{2012A&A...546A..50K}, where the authors reported longer periods travelling to larger distances even in polar regions. However, it was later noted that this behaviour was due to the availability of higher power at longer periods near the base of the corona. 

\cite{2014A&A...568A..96G} also studied the damping behaviour of slow waves in polar regions. Their results indicate the presence of two different damping regimes, the stronger damping near the limb ($<10$ Mm height) and weaker damping at larger heights. Furthermore, in order to understand the frequency dependence, damping lengths were measured over three predefined period bands i.e., 4$-$6 min, 6$-$15 min, and 16$-$45 min. No preferable dependence was found near the limb but at larger heights the damping appears to be stronger in the shorter period band which the authors explain as evidence for damping due to thermal conduction. However, it needs to be pointed out that sampling the oscillations at a larger number of periods as obtained in \cite{2014ApJ...789..118K} would be more suitable and better for understanding the period dependence.

\cite{2016ApJ...820...13M} performed 3D MHD simulations based on a uniform cylinder model to investigate the damping of slow waves at multiple periods. By perturbing the pressure at one end of the loop, slow waves are driven at 4 preselected periods. The other boundary is kept open allowing the waves to propagate and escape the domain. Thermal conduction is incorporated into the model as the only damping mechanism for the waves. The gravitational stratification and the field divergence are ignored. The obtained numerical results are forward modelled to aid a direct comparison with the observations. It is found that the damping lengths are dependent on the oscillation period with a power-law index near 1 agreeing with the previous observations for active region loops \citep{2014ApJ...789..118K}. The authors also derived full solutions to the dispersion relation for thermal conduction damping, instead with the approximations made in \cite{2014ApJ...789..118K}, which shows a similar dependence. Based on this, the discrepancy in the previous results was explained as due to the differences in the damping of shorter and longer periods which conform to strong and weak thermal conduction limits, respectively. However, the negative dependence found in the polar regions remains unexplained. In order to verify and examine if the frequency dependence obtained in the polar regions is general, \cite{2018ApJ...853..134M} performed a statistical study by employing 62 datasets of polar coronal hole regions observed between 2010 and 2017 by SDO/AIA.
\begin{figure}
\includegraphics[width=0.98\textwidth]{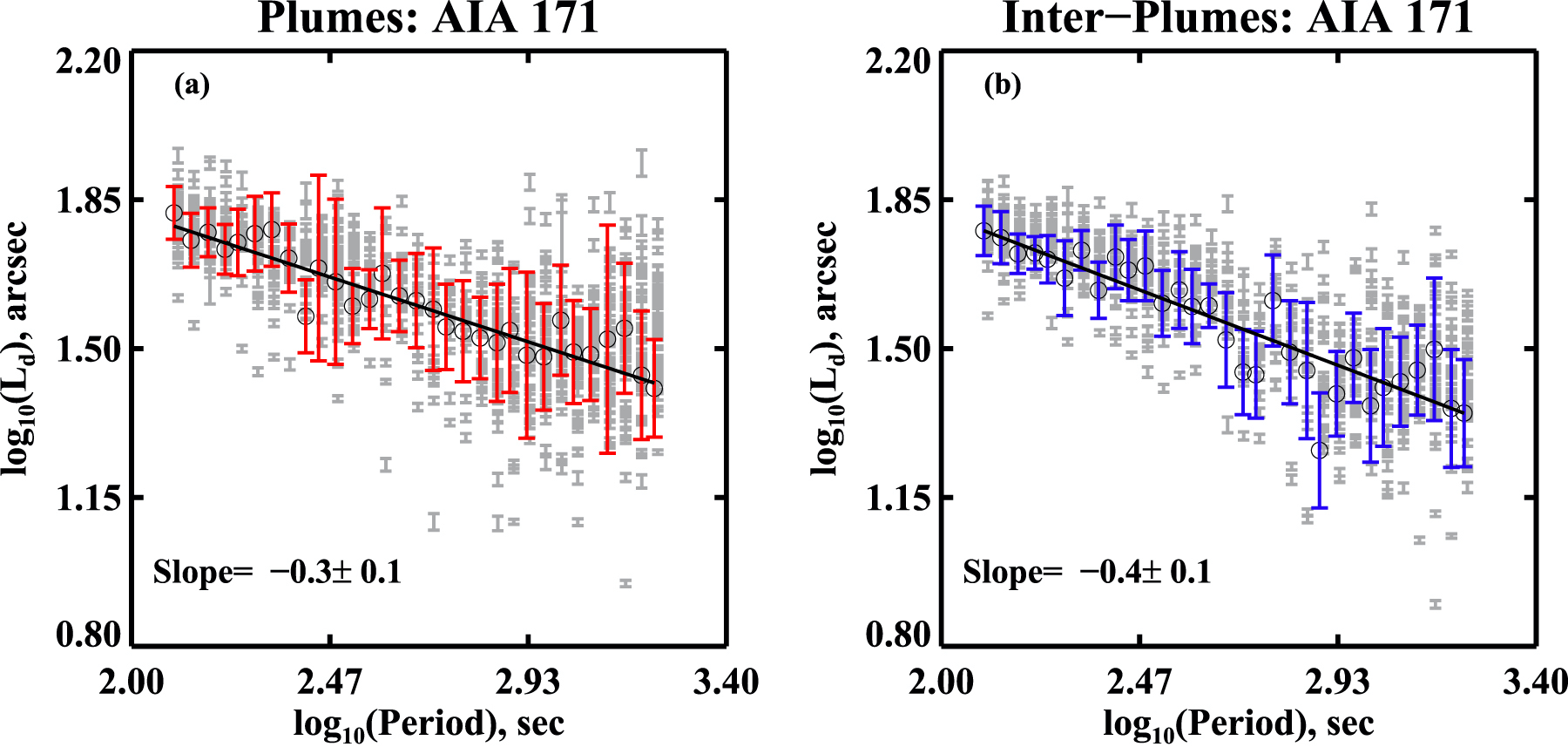}
\caption{The dependence of damping length on oscillation period extracted from the polar plume (left) and interplume (right) regions using data from the AIA 171{\,}{\AA} channel. The symbols in grey correspond to the individual measurements. The most frequent value (modal value) at each period bin is denoted by an open black circle and the standard width of the associated distribution is shown as an error bar. The slanted black line over the data represents the best fit to the most frequent values. The slope obtained from the fit is listed in the individual panels. From \cite{2018ApJ...853..134M}}
\label{kpfig3}
\end{figure}
Fig.~\ref{kpfig3} displays the results of this study for the data from the AIA 171{\,}{\AA} channel. The left and right panels show the dependences obtained for the plume and interplume regions respectively. The measured slopes obtained by fitting the modal values at each period bin are $-$0.3$\pm$0.1 for plume and $-$0.4$\pm$0.1 for interplume regions. Similar values were obtained for the data from the AIA 193{\,}{\AA} channel. These results not only confirm those from \cite{2014ApJ...789..118K}, that the anomalous dependence of damping length on oscillation period observed in polar regions is real, but also suggest that such a damping relation is typical for slow waves in polar structures which therefore, strongly necessitates an explanation.

More recently, \cite{2019FrASS...6...57S} studied the dependence of damping length on the temperature of the loop. The damping lengths of oscillations were measured in 35 loop structures (selected from 30 different active regions) and the corresponding plasma temperatures were obtained from Differential Emission Measure (DEM) analysis \citep{2012A&A...539A.146H}.
\begin{figure}
\includegraphics[width=0.98\textwidth]{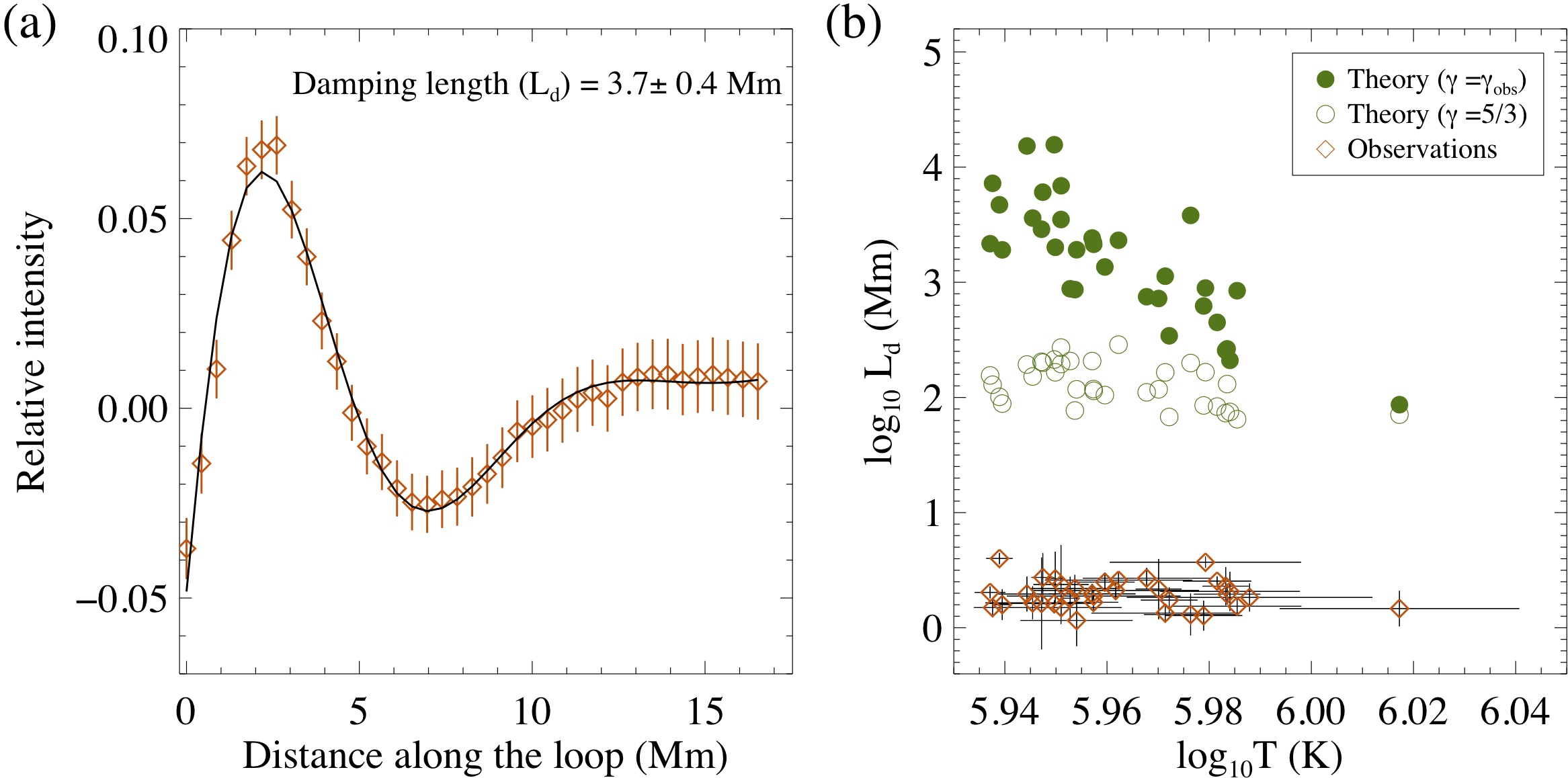}
\caption{\textit{Left}: Relative intensity of oscillation as a function of distance along a sample loop. The orange diamond symbols denote the actual data with the corresponding uncertainties shown as error bars. The solid black curve represents the best fit for an exponentially decaying sine wave model. The damping length value obtained from the fit is listed in the figure. \textit{Right}: Dependence of damping length on loop temperature. The orange diamond symbols denote the values from observations with the corresponding uncertainties shown as horizontal and vertical error bars. The filled green circles highlight the expected values based on a linear slow wave theory incorporating  damping due to thermal conduction. The open green circles denote the same but for a constant value (5/3) for the polytropic index. From \cite{2019FrASS...6...57S}}
\label{kpfig4}
\end{figure}
Fig.~\ref{kpfig4} displays the obtained results from this study. The left panel demonstrates the measurement of damping length ($L_d$) by following the phase of a sample oscillation with the obtained $L_d$ value listed in the plot. An alternate method to estimate damping length, by directly tracking the oscillation amplitudes along the loops produces similar results. The right panel of Fig.~\ref{kpfig4} depicts the obtained dependence of damping length on temperature. The expected values from linear wave theory considering thermal conduction as the damping mechanism are also plotted (green filled circles). The oscillation periods, densities, temperatures, and polytropic indices for the individual loop structures, extracted previously by \cite{2018ApJ...868..149K}, are incorporated into these calculations. Additionally, the expected values derived by assuming a constant value of 5/3 (adiabatic index expected in the corona) for the polytropic index are plotted (green open circles). As may be seen, the damping length is expected to decrease evidently to lower values for hotter loops but the observations appear to defy this. Based on these results, the authors concluded that thermal conduction is suppressed in hotter loops which is consistent with previous reports by \cite{2015ApJ...811L..13W}. Furthermore, it may be noted that the theoretical damping length values are 2$-$3 orders of magnitude larger than the observed values. While this discrepancy could be partly explained by the fact that the observed damping length values are projected values (meaning they are lower limits), the authors speculate that such a large difference could alternatively imply that thermal conduction is not the dominant damping mechanism in those coronal loops. 

Another aspect that has become popular in recent years is the misbalance in the local thermal equilibrium caused by slow waves which in turn could affect the wave dynamics significantly \citep{2017ApJ...849...62N}. To study this effect, the heating function, $H$, is parametrised as a function of the equilibrium parameters density $\rho$ and temperature $T$, and included in the energy equation of MHD in the recent models \citep{2019A&A...628A.133K, 2019PhPl...26h2113Z, 2020A&A...644A..33K, 2021SoPh..296...20P}. In comparison, in earlier models, $H(\rho,T)$ was either ignored or assumed to be a constant balanced by the other dissipative effects (e.g., radiative cooling). During the linearisation process in the modelling of waves, the partial derivatives of the heating function $\partial H/\partial \rho$ and $\partial H/\partial T$ play a crucial role, introducing new effects on the wave dynamics. In particular, \citet{2019A&A...628A.133K} found that these novel effects may result in additional damping of slow waves, even in the complete absence of thermal conduction. This could also be an interesting avenue to pursue the explanation of earlier results with unexpected dependence of slow wave damping on oscillation period or temperature \citep{2014ApJ...789..118K, 2019FrASS...6...57S}. Indeed, it seems quite promising, because the variation of the polytropic index with loop temperature \citep{2018ApJ...868..149K} is also recovered in the model \citep{2019PhPl...26h2113Z}. \\
 
\subsection{Seismological applications}
One of the important applications of studying waves in the solar atmosphere is that it gives us the ability to infer important physical parameters through seismology. The common occurrence of slow magnetoacoustic waves makes them even more compelling to pursue this. Over the past decade a number of important parameters including the plasma temperature, coronal loop geometry, coronal magnetic field, thermal conduction coefficient, polytropic index, have been extracted from the observations of slow waves in combination with MHD theories demonstrating their vast potential in this area \citep[for e.g.,][]{2009ApJ...697.1674M, 2009A&A...503L..25W, 2011ApJ...727L..32V, 2014A&A...561A..19Y, 2015ApJ...811L..13W, 2016NatPh..12..179J, 2017ApJ...834..103K, 2018SoPh..293....2D, 2018ApJ...868..149K, 2019ApJ...886....2W, 2020A&A...644A..33K}. A subset of these applications, particularly those reported in the recent years, will be discussed here in detail.

Considering a cylindrical flux tube geometry for coronal loops, within the long-wavelength limit, i.e., when the wavelength of the oscillation, $\lambda$, is much greater than the diameter of the loop, $d$, the slow waves propagate at tube speed, $c_t$, which is dependent on both the sound speed, $c_s$, and the Alfv\'{e}n speed, $v_a$, through the relation $c_t^2 = c_s^2 v_a^2/(c_s^2+v_a^2)$. From the observations of slow waves we know that the long-wavelength limit is generally applicable in the solar corona which means unless the plasma $\beta$ is very small, the propagation speed of slow waves is not only dependent on the plasma temperature but also on the magnetic field strength. By applying this relation to the observations of standing slow waves, \citet{2007ApJ...656..598W} have earlier demonstrated that one could derive the coronal magnetic field strength. More recently, \citet{2016NatPh..12..179J} have further exploited this relation by extending it to the propagating slow waves to derive spatially resolved coronal magnetic fields over a sunspot. \citet{2016NatPh..12..179J} have shown that the propagating slow waves observed in an isolated sunspot from the active region NOAA 11366 possess multiple periodicities with longer periods dominant at farther distances from the sunspot centre. The apparent propagation speeds of longer period waves were found to be significantly smaller compared to that of the shorter periods found near the sunspot centre. By combining the speed-period and period-distance dependences, the authors derived a radial dependence of apparent propagation speed on distance. In order to estimate the real propagation speed the inclination of the magnetic field needs to be known. In principle, the local magnetic field geometry can be obtained from stereoscopic observations of propagating slow waves \citep{2009ApJ...697.1674M, 2020ARA&A..58..441N} or magnetic field extrapolation methods. Although, each of these methods have their own intrinsic difficulties and ambiguities, the authors applied both the methods to derive the inclination of the field incorporating which they deduced the actual propagation speed ($c_t$) as a function of distance from the sunspot centre. The corresponding temperatures and densities were estimated employing a differential emission measure technique. Inputting these values in the expression for $c_t$, \citet{2016NatPh..12..179J} have obtained a magnetic field strength of 32$\pm$5 G near the sunspot centre which is found to rapidly decrease to about 1 G over a distance of 7 Mm. It is not explicitly stated whether these values are consistent with those derived from the extrapolations. Also, in order to improve the signal, the authors had to average the observations in the azimuthal direction so they could only provide an average radial dependence of magnetic field strength but not a fully resolved spatial distribution. Furthermore, the azimuthal averaging would result in the loss of some important structure in this dimension \citep{2014A&A...569A..72S, 2017ApJ...842...59J, 2019ApJ...877L...9K} so this is perhaps not an ideal procedure. Nevertheless, this study clearly highlights the scope of slow waves in determining one of the most important physical parameters of the corona, the magnetic field strength.

Under the condition of low-plasma $\beta$, which is generally assumed for solar corona, the tube speed $c_t$ is almost equivalent to the sound speed $c_s$. So it is common to associate the propagation speed of slow waves with the local sound speed which, in turn, is proportional to the square root of the corresponding plasma temperature, $T$. Thus, one could derive the temperature of the plasma by measuring the propagation speed of slow waves. However, the observations are typically constituted of two-dimensional projected images of the oscillations and hence could only provide a lower-limit on the slow-wave speed. Availing a rare opportunity, \citet{2009ApJ...697.1674M} analysed the propagation of slow waves in a coronal loop observed by two spacecrafts from two different vantage points. These stereoscopic observations made by \textit{STEREO}/EUVI enabled the authors to study the three-dimensional propagation of slow waves and provide their true propagation speed. The obtained propagation speed was about 132$\pm$11{\,}km{\,}s$^{-1}$ corresponding to a plasma temperature of 0.84$\pm$0.15 MK derived under the adiabatic assumption. The observed propagation speeds are, in general, constant along the length of the loop. Earlier observations by \citet{2003A&A...404L...1K} report differential propagation of slow waves in different temperature channels indicating a sub-resolution structure within a coronal loop. However, these authors also did not find any spatial variation of propagation speeds in the individual channels.
\begin{figure}
\includegraphics[width=0.31\textwidth]{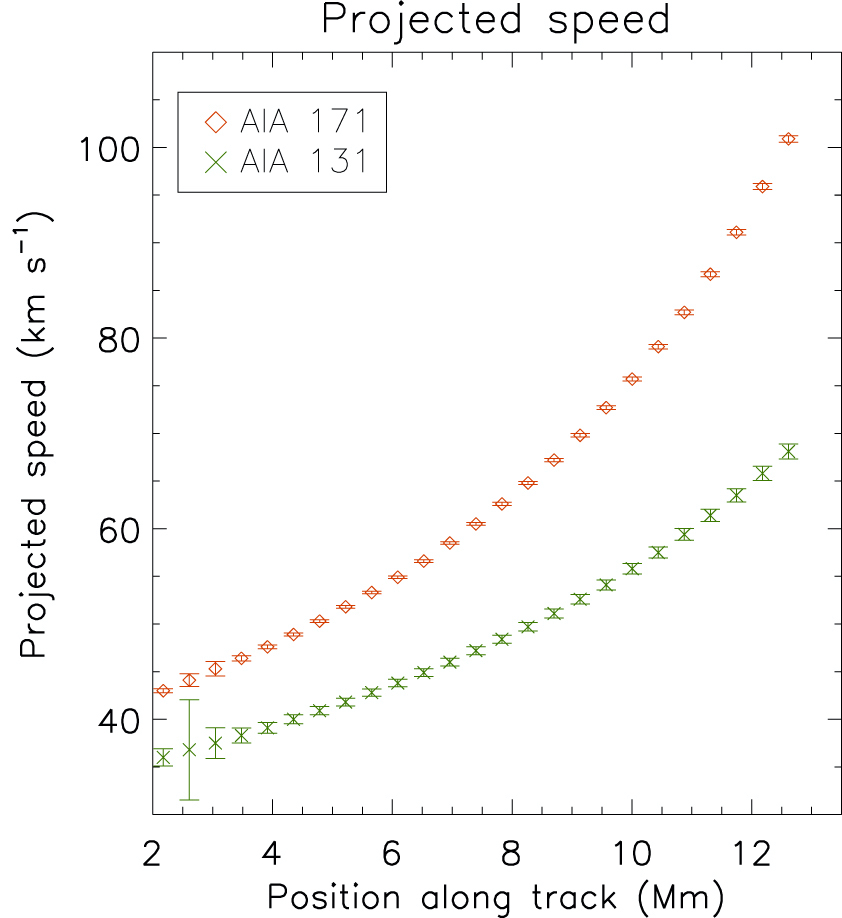}
\includegraphics[width=0.67\textwidth]{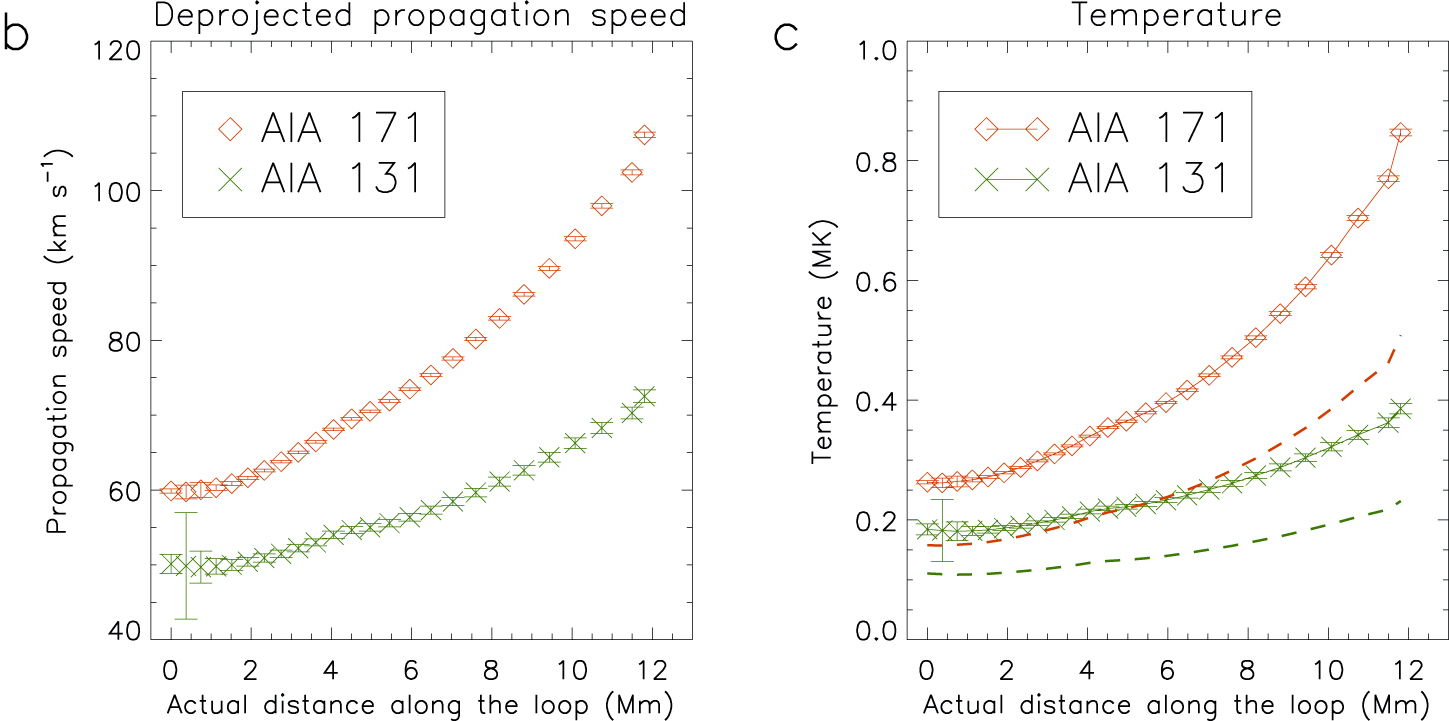}
\caption{The propagation speed of slow waves measured along a selected loop track (left), corresponding deprojected values (middle) and the derived plasma temperatures (right). Different symbols/colors represent results for different channels as described in the figure legend. Note the distances in middle and right panels correspond to deprojected values from the bottom of the loop. The diamond and `x' symbols in the right panel denote the plasma temperatures derived under the isothermal approximation. The dashed lines show corresponding values for an adiabatic case. Adapted from \citet{2017ApJ...834..103K}.}
\label{kpfig5}
\end{figure}
Recently, using the high-spatial resolution observations from \textit{SDO}/AIA \citet{2017ApJ...834..103K} studied the propagation of slow waves along an active region fan loop in two different temperature channels. In addition to the differential propagation, the authors found an accelerated propagation of waves along the loop in both the channels. This enabled them to derive the spatial variation of plasma temperature along the wave propagation path. The authors employed nonlinear force-free magnetic field extrapolations to estimate the inclination of the loop and thereby deduced the deprojected propagation speeds. Fig.{\,}\ref{kpfig5} shows the initial observed speeds, deprojected true speeds, and  the obtained temperature values as a function of distance in the left, middle, and right panels, respectively. The results for both wavelength channels are shown in the figure. The temperatures were estimated assuming isothermal propagation of slow waves although corresponding values for adiabatic propagation are also shown (dashed lines in the right panel). It may be noted that the temperature values especially near the bottom of the loop ($\approx$0.2{\,}MK) are much lower compared to the previous results. This is mainly because of the lower propagation speeds observed here. It is possible that there are aperiodic downflows in the loop concurrent with the waves. Although the time-distance maps do not clearly indicate this, such flows are fairly common in sunspots \citep{2015A&A...582A.116S, 2016A&A...587A..20C, 2018ApJ...859..158S, 2020A&A...636A..35N, 2020A&A...640A.120N}. If that is the case, the true propagation speed of the wave (and hence the plasma temperature) could be higher. Nevertheless, the evidence for differential propagation is remarkably clear suggesting a multi-thermal structure within the coronal loop. The authors highlight this as an ability of slow waves to resolve the sub-resolution structure of coronal loops.

Utilising the spectroscopic observations of running slow waves in a coronal loop made by \textit{Hinode}/EIS, \citet{2011ApJ...727L..32V} derived the effective adiabatic index or the polytropic index ($\gamma_\mathrm{eff}$) of coronal plasma. The authors employed the expected relation between the density and temperature perturbations due to a slow wave to achieve this. The required density and temperature estimations were made from appropriate spectroscopic line ratios. The authors obtained a value of $\gamma_\mathrm{eff} =$1.10$\pm$0.02, close to the isothermal value 1, implying that thermal conduction is very efficient in the solar corona. Furthermore, using the observed phase lag between the density and temperature perturbations (under the assumption that it is entirely introduced by thermal conduction), they derived the thermal conduction coefficient, $\kappa_0 =$9$\times$10$^{-11}${\,}W{\,}m$^{-1}${\,}K$^{-1}$, which is of the same order of magnitude as the classical \textit{Spitzer} conductivity. On the other hand, a similar study by \citet{2015ApJ...811L..13W} using the observations of standing slow waves in hot coronal loops, resulted in a polytropic index value of 1.64$\pm$0.08. The proximity of this value to the adiabatic index of 5/3 (for an ideal monatomic gas) implies that thermal conduction is heavily suppressed in the investigated loop. By comparing the observed phase lag between density and temperature perturbations to that of a 1D linear MHD model, the authors suggest that thermal conduction is suppressed at least by a factor of three. \citet{2018ApJ...868..149K} investigated the behaviour of propagating slow waves in multiple fan loops from 30 different active regions. By analysing the relation between corresponding density and temperature perturbations these authors also derived associated polytropic indices which were found to vary from 1.04$\pm$0.01 to 1.58$\pm$0.12. Additionally, the authors studied their temperature dependence which indicates a higher polytropic index value for hotter loops. It was further concluded that this dependence perhaps suggests a gradual suppression of thermal conduction with increase in temperature of the loop thus bringing both the previous studies into agreement.

Coming back to the new models with the thermal misbalance \citep[see e.g.][and references therein]{2019A&A...628A.133K, 2019PhPl...26h2113Z}, these offer an excellent opportunity for slow waves to probe the coronal heating function, which is arguably the greatest mystery in solar or stellar physics. In order to study this \citet{2020A&A...644A..33K} has parameterised the heating function $H$ as power laws of density $\rho$ and temperature $T$. Based on the common existence of long-lived plasma structures and the statistical properties of the slow wave damping, the authors then constrained the power-law indices of these key quantities offering an insight into the characteristics of coronal heating function. It has also been shown that the characteristic thermal misbalance time scales are on the same order as the oscillation periods and damping times of slow waves observed in various structures including flaring loops and coronal plumes. Although this investigation has been largely theoretical, it offers a lot of observational potential to make progress in the decades old question of characterising the coronal heating mechanism. For more details, please see the review by \citet{2020SSRv..216..140V} in this volume. \\

\section{Alfv\'{e}nic/Kink waves}
\label{alfven_waves}

Apart from slow magnetoacoustic waves, kink(Alfv\'enic) waves are also observed in the solar atmosphere at the locations of the open magnetic field regions. Alfv\'enic waves were suggested to have mixed properties as they propagate, having both compression and parallel vorticity and with non-zero radial, perpendicular and parallel components of displacement and vorticity \citep{2009A&A...503..213G,2012ApJ...753..111G,2019FrASS...6...20G,2021A&A...646A..86G}. Thus kink waves under some physical conditions, could share properties of surface Alfv\'en waves \citep[Also see][]{2009A&A...503..213G}.
In this review we will use terms Alfv\'enic and kink interchangeably.

\subsection{Observations}
 Both spectroscopic and imaging observations have been used to establish the presence of the Alfv\'enic waves in the solar atmosphere. In the following subsections we discuss them separately. 

\subsubsection{Spectroscopic Observations}
\label{section:Spectroscopic Observations}
\begin{figure}[ht!]
\centering
\includegraphics[scale=0.25]{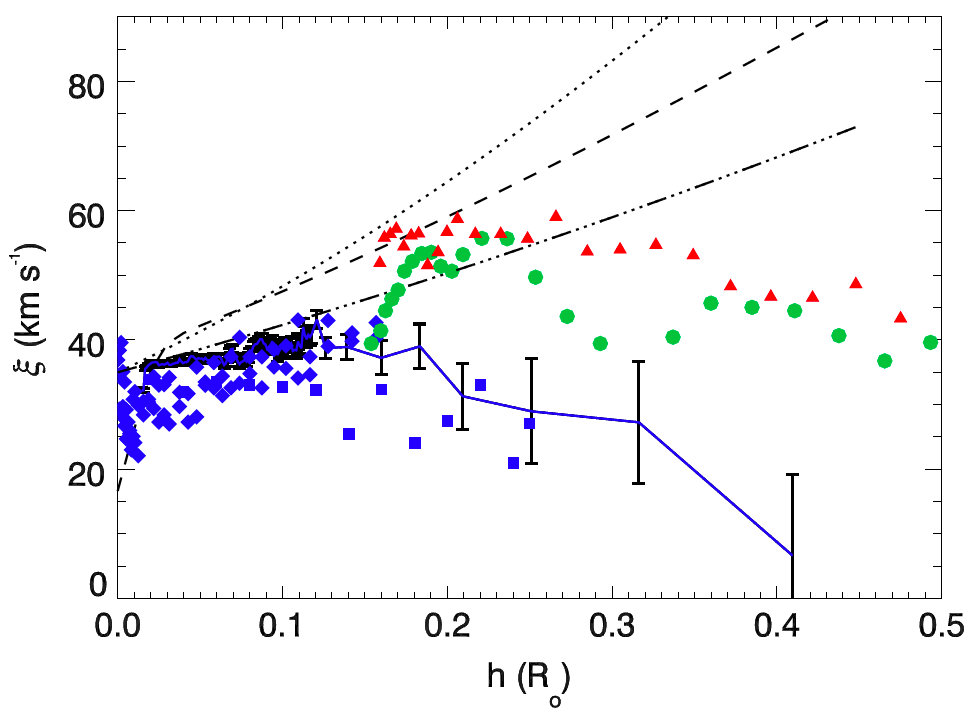}
\caption{Variation of the nonthermal line widths with height above a polar coronal hole in different spectral lines. Red and green color represents the nonthermal line widths derived from the eclipse observations in Fe XIII and Fe XIV emission lines, respectively \citep{2011SoPh..270..213S}. Square and diamond in blue are derived from the observations taken in Fe XII using SUMER and EIS, respectively \citep[see,][respectively]{2003ApJ...598..657M,2009A&A...501L..15B}. Solid line represents the nonthermal line widths derived from Fe XII emission line \citep{2012ApJ...751..110B}. The dotted, dashed and dash-dotted curves represent the nonthermal widths expected from WKB propagating for different density profiles as reported in \citet{1999JGR...104.9801G,2012ApJ...751..110B,1999A&A...349..956D}, respectively.  Adapted from \citet{2012ApJ...751..110B}}.
\label{ntlw}
\end{figure}

Perhaps one of the earliest signatures of the presence of Alfv\'enic and Alfv\'en waves (or turbulence) in the solar atmosphere comes from the observations of the nonthermal spectral line broadenings in the solar atmosphere. \citep{1973ApJ...181..547H,2008ApJ...676L..73V,1976ApJ...205L.177D, 1976ApJS...31..417D, 1976ApJS...31..445F, 1990ApJ...348L..77H}. The nonthermal line broadening at the location of the coronal holes was reported by \citet{1998A&A...339..208B, 1998SoPh..181...91D, 2005A&A...436L..35O,1997ApJ...476L..51O,1999ApJ...510L..59K,2016SSRv..201...55A} using the Solar Ultraviolet Measurements of Emitted Radiation (SUMER) spectrometer, Coronal Diagnostic Spectrometer (CDS), and Ultraviolet Coronagraph Spectrometer (UVCS) on-board {\it Solar and Heliospheric Observatory} (SoHO). Furthermore, an increase followed by the flattening of the nonthermal line widths with height above the solar limb was noted \citep{1998A&A...339..208B, 1998SoPh..181...91D, 2005A&A...436L..35O,1990ApJ...348L..77H}. A similar nature of variation of the nonthermal line widths was found at the location of the coronal holes using Extreme Ultraviolet Imaging Spectrometer (EIS) on-board {\it Hinode}  \citep{2007ApJ...667L.109D,2009A&A...501L..15B,2012ApJ...753...36H}. Using the nonthermal line broadening these studies have estimated the energy flux carried by the Alfv\'enic waves in the solar atmosphere and found that it is enough to balance total coronal energy losses in coronal holes. Furthermore the flattening or leveling-off of the nonthermal line widths was attributed to Alfv\'en wave damping (see Figure~\ref{ntlw}). The departure from the expected WKB theory (undamped propagation of Alfv\'en wave) happens as early as 0.1 $R_{\odot}$ from the solar surface.

However, recently \citet{2020ApJ...899....1P} have shown analytically and using numerical simulations that for a period averaged Alv\'en(ic) waves (such as those observed by SUMER with large exposure time), the nonthermal line widths are larger than the rms wave amplitude by a factor of $\sqrt{2}$. Further, LOS integration of several structures oscillating with different polarisation and phases of oscillation cause nonthermal line widths of the order of rms wave amplitudes. These results are in contrast with those obtained in some of the above mentioned studies \citep[ e.g,][]{1990ApJ...348L..77H,1998A&A...339..208B, 1998SoPh..181...91D, 2005A&A...436L..35O,1997ApJ...476L..51O,1999ApJ...510L..59K}, where authors assume that the nonthermal line widths are smaller than the rms wave amplitudes by a factor of $\sqrt{2}$. Thus the energy flux, which varies as the square of the rms wave amplitudes, is overestimated at least by a factor of 2 in above mentioned studies.\\
Additionally, several analytical  and numerical Alfv\'en wave models  have also been made to demonstrate the propagation and dissipation of the Alfv\'en waves and their potential contributions in the observed line-width variations in the solar atmosphere, thus invoking the significant role of these waves in generating turbulence and/or heating  \citep[e.g,][and references therein]{2002MNRAS.336.1195P,2006SoPh..237..143D,2013MNRAS.428...40C,2015ApJ...811...88Z,2015A&A...581A.131J}. However, such studies do not consider the present day scenario of the mixed wave modes/Alfv\'enic wave modes and their specific physical properties and role in the line-width broadening or narrowing.\\

\begin{figure}[ht!]
\centering
\includegraphics[scale=0.35]{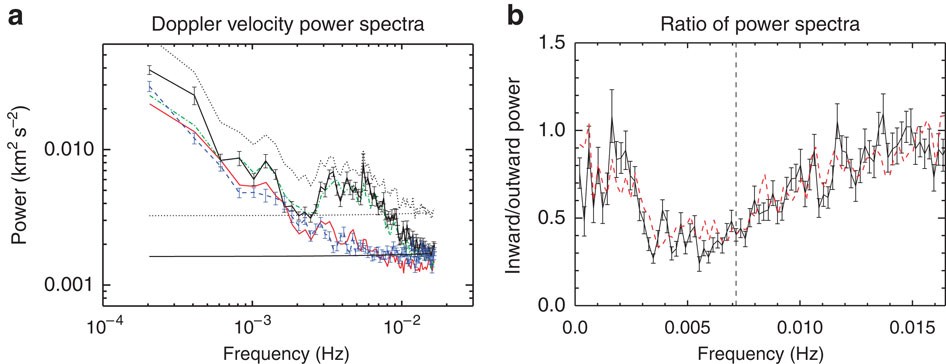}
\caption{{\it (a)} Power spectra of the Doppler velocities showing enhancements at 3-5 mHz. {\it (b)}: Frequency dependent ratio of inward and outward power suggesting that low frequency waves are more likely to reflect \citep[adapted from][] {2015NatCo...6E7813M}. }
\label{ps}
\end{figure}

The Coronal Multi-channel Polarimeter (CoMP, see \citealt{2007Sci...317.1192T}) has established the ubiquity of Alfv\'enic/kink waves in the solar corona, primarily observed as  propagating Doppler velocity fluctuations. \citet{2009ApJ...697.1384T} reported these fluctuations have both an outwardly- and inwardly-propagating nature and a disparity between the outward and inward wave power, which could be a signal of isotropic MHD turbulence. Furthermore, \citet{2015NatCo...6E7813M} observed Alfv\'enic waves that partially reflect while propagating up along the open magnetic field regions, such as polar coronal holes. These results demonstrate that counter-propagating Alfv\'enic waves exist in open coronal magnetic fields, thus providing support for Alfv\'en wave turbulence models (see also section \ref{section:Generation of turbulence in open magnetic field regions}). Interestingly, there is also a disparity in outward-to-inward wave power for closed magnetic fields (see \citealt{2010A&A...524A..23T, 2010ApJ...718L.102V, 2019ApJ...876..106T}). Moreover, low frequency Alfv\'enic waves are more likely to reflect than high frequencies (see the right panel of Figure~\ref{ps}). Earlier, theoretical studies have also reported a similar behaviour for Alfv\'en waves \citep{2005ApJS..156..265C}. These authors reported that these waves are strongly reflected at the transition region (95\%). While beyond transition region, waves with frequency lower than the critical frequency (local gradient of Alfv\'en velocity) reflect, which is about 30 hours ($\sim$1 day) at large distances from Sun. Again using CoMP, an enhancement of the power spectra in 3-5 mHz was noted in the open magnetic field regions (see Figure~\ref{ps}), suggesting the possible role of p-modes in generating Alfv\'enic waves, as a spatially-ubiquitous input  \citep{2016ApJ...828...89M} that is sustained over the solar cycle \citep{2019NatAs...3..223M}. An enhancement in power at 3-5 mHz is also seen in the intensity variations from SDO/AIA in quiet sun region and sunspots \citep{2016A&A...592A.153K}.

\begin{figure}[ht!]
\centering
\includegraphics[scale=0.75]{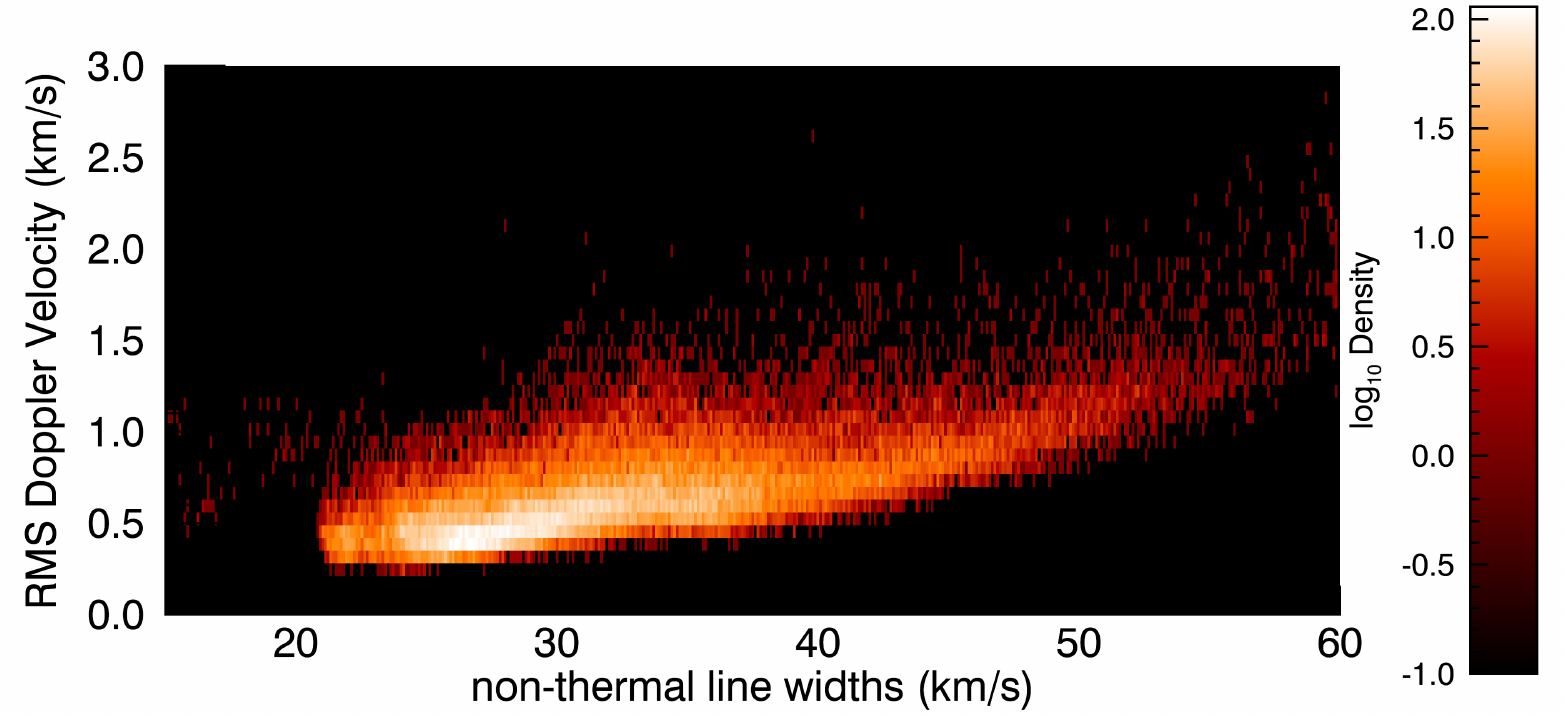}
\includegraphics[scale=0.55]{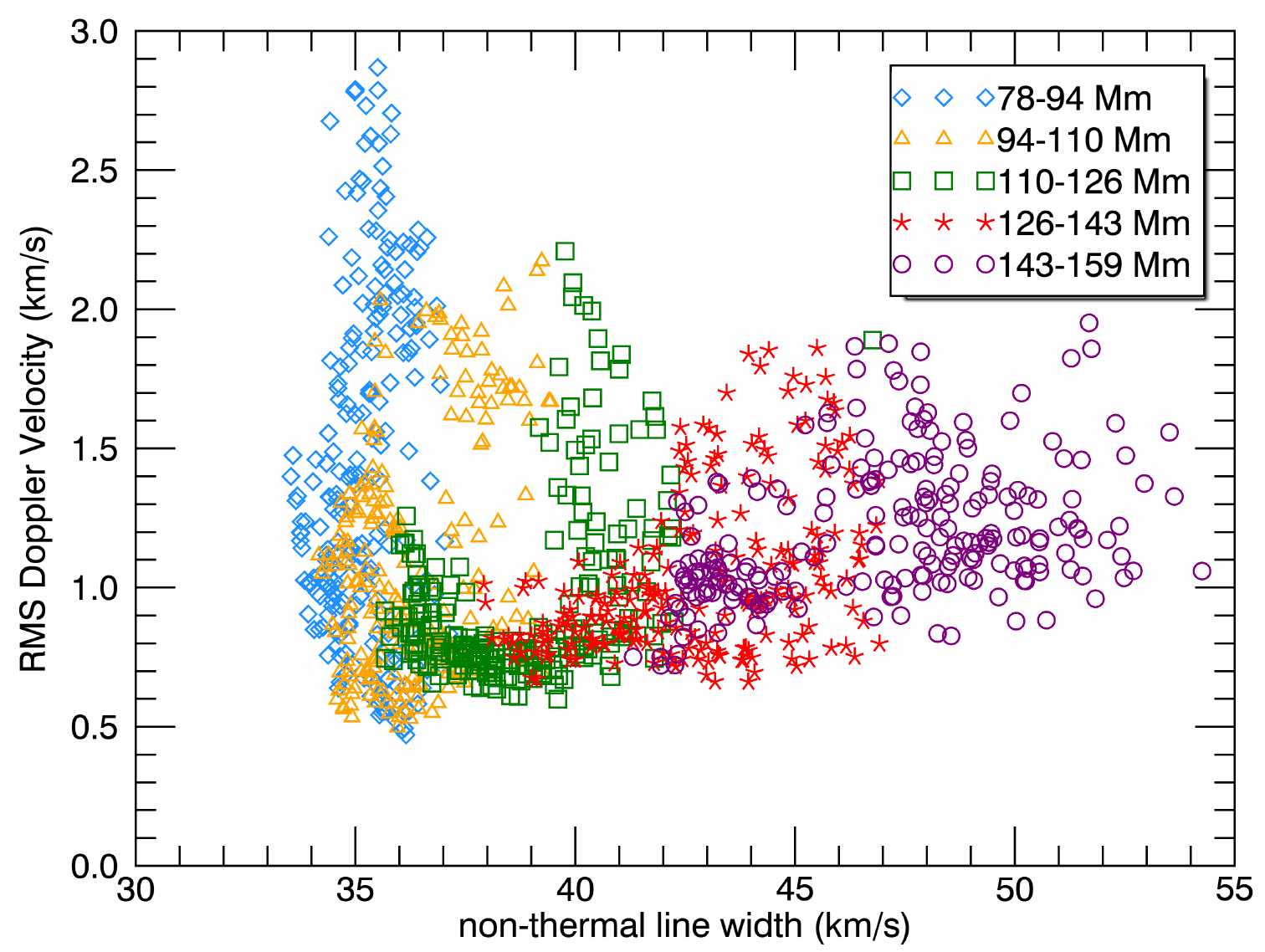}
\caption{{\it Top} Wedge-shaped correlation displaying the distribution of the rms Doppler velocities and mean nonthermal line widths over entire CoMP FOV. {\it Bottom}: At the location of coronal holes segregated with height. Adapted from \citet{2019ApJ...881...95P}.}
\label{wedge}
\end{figure}

Furthermore, \citet{2012ApJ...761..138M} have noted a wedge shape correlation between rms Doppler velocities and nonthermal line widths in the solar corona similar to Figure~\ref{wedge}. Using Monte Carlo method, these authors suggested that the observed correlation might be due to the Alfve\'enic wave propagation and LOS integration of several oscillating structures. Since the amplitude of Alfv\'enic wave increases with height, LOS integration of several oscillating structures also increases non-thermal line widths. This lead to a wedge shape correlation. However, they artificially added a substantial amount of nonthermal broadening in the Monte Carlo simulations to match the observed nonthermal broadening. The source of this additional nonthermal bradening was not known. Later, \citet{2019ApJ...881...95P} have studied the correlation between rms Doppler velocities and the nonthermal line widths at different heights at the location of the polar coronal holes using observations from CoMP (see bottom panel of Figure~\ref{wedge}) and numerical simulations. Combining numerical simulations and forward modelling, these authors could reproduce the observed wedge-shaped correlation without adding any artificial source of nonthermal line widths (see left panel of Figure~\ref{fig-wedge}). They proposed that at least a part of the wedge-shape correlation between the nonthermal line widths and rms Doppler velocities is due to the propagation of Alfv\'enic waves in the solar atmosphere and a part is due to height dependence of RMS wave amplitude and non-thermal line widths. (see section \ref{section:forward modelling} for more details). Figure~\ref{wedge} denotes a wedge shape correlation over entire solar corona (top panel) and at the location of the coronal holes (bottom panel) at different heights obtained using the data from CoMP. 

\subsubsection{Imaging Observations}
\label{section:Imaging Observations}


\begin{figure}[t]
\centering
\includegraphics[scale=0.7]{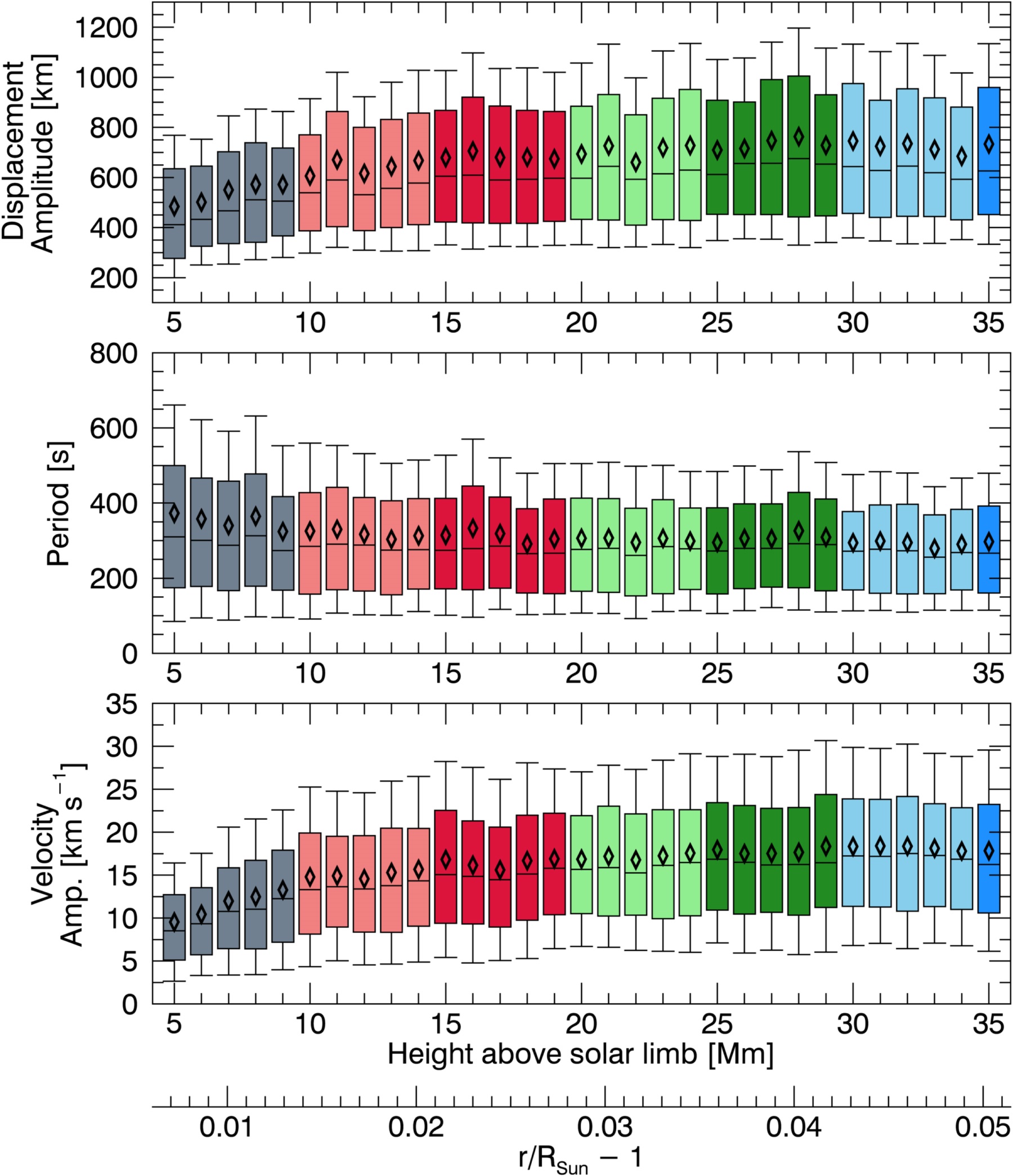}
\caption{Box and whisker plot of wave parameters for slits analysed (31 in total). Lower and upper bounds of each coloured box correspond to the first and third data quartiles, respectively. Horizontal lines indicate median values, diamonds denote log-normal means, \lq{whiskers}\rq{} show the log-normal standard deviations. Solely as a visual aide, every five slits appear in the same colour. \citep[adapted from][] {2020ApJ...894...79W}. }
\label{Figure: Weberg et al 2020}
\end{figure}

In addition to the nonthermal broadening, direct observation of transverse waves have been reported in  polar coronal holes. Using data from SDO/AIA, \cite{2011Natur.475..477M} observed ubiquitous outward-propagating Alfv\'enic motions in the corona with amplitudes of the order of 25 $\pm$ 5 km s$^{-1}$, uniformly-distributed over periods of 150-550 seconds throughout the quiescent solar atmosphere. These estimates were obtained using Monte Carlo simulations, which we refer to as indirect measurements. Using these indirect measurements, the energy flux carried by these waves was estimated to be enough to accelerate the fast solar wind and power the quiet corona. Again using SDO/AIA, \citet{2014ApJ...790L...2T}  provided the first {\it{direct}} measurements of transverse wave motions in solar polar plumes but, in contrast to \cite{2011Natur.475..477M}, found lower Alfv\'enic wave amplitudes around 14 $\pm$ 10 km s$^{-1}$, as well as observing a broad distribution of parameters which is skewed in favour of relatively-smaller amplitude waves. When this whole range of parameters is taken into account to calculate the energy flux carried by these waves, the energy budget, determined by direct measurement, falls 4-10 times below the minimum theoretical requirement (models require 100–200 W m$^{-2}$ of wave energy flux near the coronal base to match empirical data, see, e.g. \citealt{1988ApJ...325..442W} and \citealt{1995JGR...10021577H}). Furthermore, these energy budget calculations assume a filling factor of unity in a homogeneous plasma, which corresponds to the most generous case in terms of energy flux. Thus, more realistic estimates about the inhomogeneity of the medium would yield even smaller energy fluxes (see, e.g. \citealt{2013ApJ...768..191G}).

\citet{2014ApJ...790L...2T} conclude that transverse waves in polar coronal holes are insufficiently energetic to be the dominant energy source of the fast solar wind. Furthermore, it is interesting to note that the directly observed transverse velocity amplitudes are  broadly compatible (within standard deviations) with the line-of-sight nonthermal velocities reported in plume plasma by \citet{2009A&A...501L..15B} using Hinode/EIS. If one assumes that these nonthermal velocity measurements contain contributions from both transverse (kink) waves and torsional Alfv\'en waves \citep{2019ApJ...870...55G}, this comparability suggests  torsional Alfv\'en waves may not contribute strongly to plume nonthermal velocity measurements (otherwise we would expect nonthermal line-of-sight velocities to be larger). However, we must be cautious since this may also be due to a lack of large-amplitude torsional motions in plumes and/or under-resolved line-of-sight motions using our current spectrometers (see also Section \ref{section:Spectroscopic Observations} for further discussion of nonthermal motions).

Recently, \citet{2020ApJ...894...79W} utilised the NUWT automated  algorithm (see \citealt{2018ApJ...852...57W} for details, as well as Section 2.2 of \lq\lq{Novel techniques in coronal seismology data analysis}\rq\rq) to analyse the transverse waves above the polar coronal hole, with a specific interest in how those properties vary with altitude. Between altitudes of 15 to 35 Mm, it was found that measured wave periods were approximately constant, and that the displacement and velocity amplitudes increased at rates consistent with undamped waves (Figure~\ref{Figure: Weberg et al 2020}). Between 5 and 15 Mm above the limb, the relative density was inferred to be larger than that expected from 1D hydrostatic models, which may indicate a more extended transition region with a gradual change in density.



\subsection{Theoretical modelling}

\subsubsection{Generation of turbulence in open magnetic field regions}
\label{section:Generation of turbulence in open magnetic field regions}

Although some observations presented in the previous subsection show that the omnipresent propagating Alfv\'enic waves have roughly enough energy to power the quiet sun, the problem of how this energy is converted to heat still persists. Because of the very high Reynolds numbers of flows in the corona 
(e.g. $R_m \approx 10^8 - 10^{12}$, see \citealt{2005psci.book.....A}), any potential damping mechanism must include the generation of scales small enough so that wave energy can be efficiently dissipated. The most popular damping mechanisms of Alfv\'en and kink(Alfv\'enic) waves are phase mixing and resonant absorption respectively.
For more on the linear damping mechansism, see, e.g., \citet{2015RSPTA.37340261A}. In this subsection we will focus instead on nonlinear damping mechanisms, such as turbulence. In MHD turbulence, damping occurs through the nonlinear cascade of wave energy down to the dissipation scales. There are several wave turbulence models which deal with coronal heating and solar wind acceleration in open coronal structures \citep[see, e.g.,][]{2009ApJ...707.1659C,2013ApJ...776..124P,2016ApJ...821..106V,2017ApJ...835...10V,2019ApJ...880L...2S,2019JPlPh..85d9009C}. While a few compressible models are able to maintain a multi-million degree corona and accelerate the solar wind through turbulent heating, it is known that turbulent heating without compressible effects appears to be insufficient in open coronal regions \citep{2016ApJ...821..106V,2019SoPh..294...65V}, even when injecting Alfv\'en waves with rms velocities on the order of the full non-thermal line broadening observed in the corona. All of the previous turbulent models rely on the linear process of reflection of outgoing waves along radial Alfv\'en speed gradients in order to generate turbulence. The presence of nonlinearly interacting counter-propagating Alfv\'en waves is thought to be a necessary condition (see subsection \ref{section:tm_solwind}). While this requirement still holds in incompressible and transversely homogeneous plasma, as employed in most of the previous models, it is now known that MHD waves affected by perpendicular structuring, such as surface Alfv\'en, kink, or Alfv\'enic waves can self-cascade nonlinearly, leading to what is referred to as uniturbulence \citep{2017NatSR...14820M,2019ApJ...882...50M}, a term which is used to differentiate this turbulence generation mechanism from the counter-propagating one. The nonlinear energy cascade rate for kink waves propagating in a cylindrical flux tube was calculated recently by \citet{2020ApJ...899..100V}. It was shown that for thin coronal strands and kink wave amplitudes that are comparable to the observed ones the energy cascade time scale can be shorter than the wave period. Uniturbulence lifts the necessity for reflected waves in turbulence generation, however it relies on the presence of structuring. Such structuring may be obtained through other mechanisms, such as the parametric decay instability (see subsection \ref{section:tm_solwind}). As there is ample evidence for coronal structuring perpendicular to the magnetic field \citep[see, e.g., ][]{2014ApJ...788..152R,2018ApJ...862...18D}, uniturbulence might play a major role in turbulence generation. Preliminary results show this is indeed the case \citep{2021ApJ...907...55M}, however more research needs to be conducted in order to estimate the added contribution to coronal heating and solar wind acceleration. 

\subsubsection{Forward modelling of Alfv\'enic waves in the open magnetic field regions}
\label{section:forward modelling}
\begin{figure}[!h]
\centering
\includegraphics[scale=0.4,angle=90]{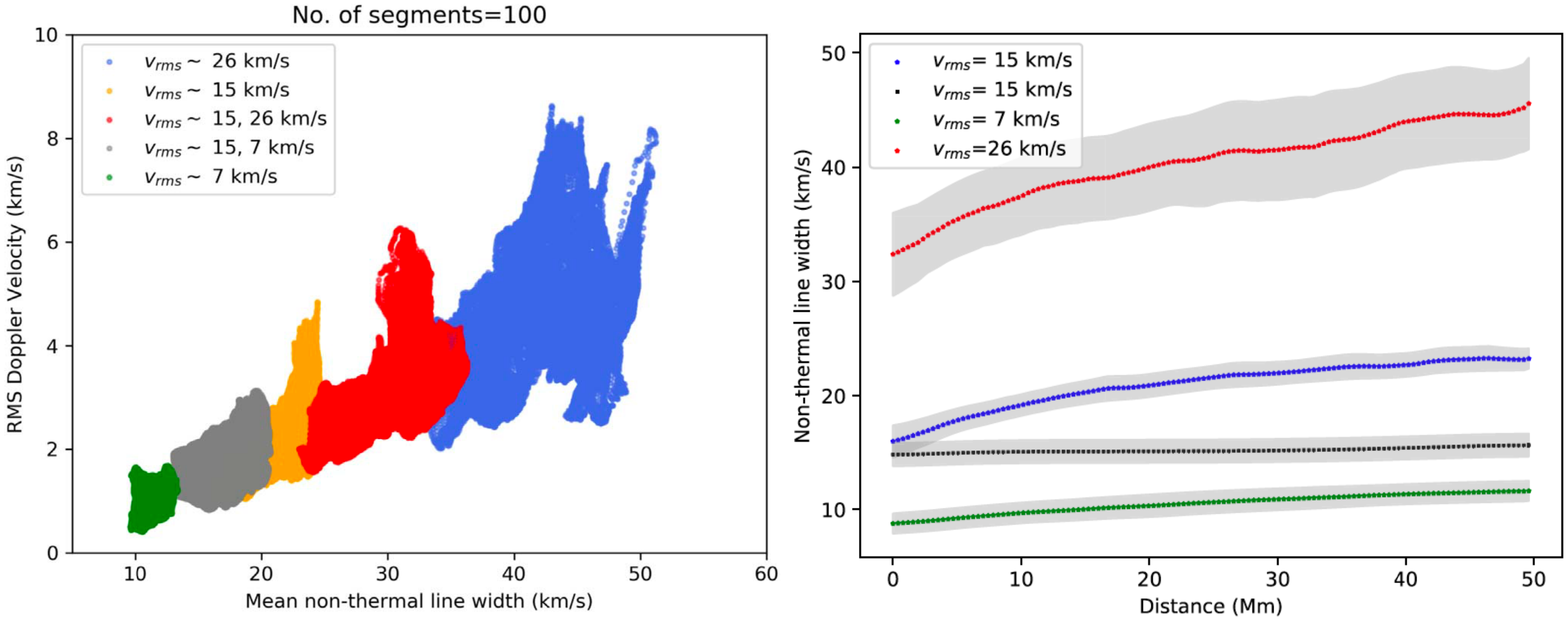}
\vspace{-2cm}
\caption{{\it Left:} Wedge-shape correlation derived from the synthetic spectra obtained by performing forward modelling on 3D MHD simulations and LOS integration over 100 segments oscillating in the random directions and in random phases. Different colors represents different rms velocities used for exciting the Alfv\'enic waves at the bottom boundary. {\it Right:} variation of the nonthermal line widths with height in the solar atmosphere. The variation qualitatively matches with those observed in the solar corona. Different colors represent different strength of the velocity drivers. Curve in black shows the variation of nonthermal line widths without gravity  \citep[adapted from][]{2019ApJ...881...95P}.}
\label{fig-wedge}
\end{figure}

The emission observed by a telescope is projected in the plane of sky and affected by the line-of-sight (LOS) effects because of the optically thin nature of the solar corona. The LOS superposition of the emission makes the interpretation of the observed spectroscopic or imaging features difficult to interpret in the solar atmosphere. To account for the LOS effects, a forward modelling of 3D numerical simulations is needed, by which the generated synthetic spectra and images can be compared with the real observations to test the validity of the MHD models \citep{10.3389/fspas.2016.00004}. \citet{2013ApJ...778..176O, 2017ApJ...845...98O} performed the forward modelling of Alfv\'en waves in the solar atmosphere incorporating both wave propagation and dissipation in open and closed magnetic field regions in optically thin EUV emission lines. A fairly good match is obtained between the total intensities and the nonthermal line widths derived from the synthetic spectra and those derived from the observations of EIS and SUMER \citep{2013ApJ...778..176O, 2017ApJ...845...98O}. Further, it was noted that the rms wave amplitudes obtained using forward modelling were smaller than those expected for the undamped waves (using WKB assumption) for both open and closed magnetic field regions. Wave dissipation due to counter-propagating waves and wave reflection are proposed to be the main mechanism for damping of Alfv\'en waves in AWSoM, causing rms wave amplitudes smaller than the expected WKB propagation \citep{2017ApJ...845...98O}. It should be noted that the perpendicular structure of Alfv\'en waves is prescribed by the driver, in contrast with collective fast magnetoacoustic modes of perpendicular plasma non-uniformities, the results of such modelling could be sensitive to the perpendicular structuring. In addition to the global model such as Alfven-Wave driven sOlar wind Model (AWSoM), models based on the collection of the single and multiple flux tubes are also used for studying the wave-driven turbulence \citep[][see section \ref{section:Generation of turbulence in open magnetic field regions} for more details.]{2017ApJ...835...10V,2017NatSR...14820M,2019ApJ...881...95P}. \citet{2012ApJ...746...31D} studied the effect of LOS integration of propagating Alfv\'enic waves in multistranded coronal loops using numerical simulations and forward modelling. They reported that the energies estimated from the derived Doppler velocities from their model are an underestimation of the actual energy present in the Alfv\'enic waves. Recently, \citet{2019ApJ...881...95P} has performed the 3D numerical simulations of propagating Alfv\'enic waves in open magnetic field regions with both transverse and longitudinal structuring leading to the reflection of waves and generation of the uniturbulence. However, it should be noted that recently \citet{2021ApJ...907...55M} have shown that the reflected wave power is quite less and the turbulence in transversely inhomogeneous plasma is primarily due to self cascading of propagating kink(Alfv\'enic) waves (uniturbulence). These authors performed the forward modelling for Fe XIII emission line centered at 10794\AA~and compared the synthetic spectra with those obtained from the CoMP. Their model was able to reproduce the observed wedge-shaped correlation between the nonthermal line widths and rms wave amplitudes. The left panel of Figure~\ref{fig-wedge} shows the wedge-shaped correlation between nonthermal line widths ans rms Doppler velocities derived from the synthetic spectra. It matches reasonably well with the correlation observed in the solar corona (see Figure~\ref{wedge}). Right panel of Figure~\ref{fig-wedge} shows the variation of nonthermal line widths for different strength of velocity drivers (shown in different colors). The nature of variation matches with those observed in the solar atmosphere using the data from SUMER \citep{1998A&A...339..208B} and EIS \citep{2012ApJ...753...36H}. The curve in black represents the variation of the nonthermal line widths with heights for a non-stratified plasma. Large nonthermal line widths in these simulations are caused by the LOS integration and uniturbulence due to transverse inhomogeneity. Comparison of the gravitationally stratified cases with the non-stratified case shows that the increase in the nonthermal line widths leading to the wedge-shaped correlation is due to the increase in the amplitude of Alfv\'enic waves with height (due to decrease in the density). Also, the Right panel of Figure~\ref{fig-wedge} shows that the nonthermal line widths levels off with height. Though, the leveling-off of the nonthermal line widths is attributed to the wave damping \citep{2012ApJ...753...36H}, \citet{2020ApJ...899....1P} have suggested that a part of leveling-off of the nonthermal line widths could be due to the non-WKB propagation of the low frequency (large wavelength) Alfv\'enic waves in the gravitationally stratified plasma. The reflection of Alfv\'enic waves is also observed in the solar corona as explained in section~\ref{section:Spectroscopic Observations}. Though the forward modelling of the uniturbulence models have shown that the LOS effects are quite important while reproducing the observed spectroscopic properties of the solar corona, the influence of uniturbulence (with minimal LOS integration) on the nonthermal line broadening is yet to be ascertained. 

\subsection{Coronal Seismology}
Apart from heating the solar corona Kink (Alfv\'enic) waves in the solar atmosphere can be used to diagnose the local plasma conditions because properties of waves depend on the properties of the medium in which they travel. Here, we will discuss the seismological application of the kink(Alfv\'enic) waves in the solar atmosphere, particularly in the open magnetic field regions. Recently, there have been a significant developments in diagnosing the plasma conditions using the principle of Bayesian inference \citep{2021ApJS..252...11A,2018AdSpR..61..655A}. \citet{2019A&A...625A..35A,2020A&A...640L..17M} have successfully used the Bayesian inference techniques to infer the local plasma parameters and field conditions for the standing and propagating kink waves. Though there have been a plethora of studies in coronal seismology dedicated to inferring the cross field density structures and magnetic fields in closed magnetic field regions, there are limited studies on the application of the coronal seismology of propagating kink waves in the open magnetic field regions. \citet{2015NatCo...6E7813M} applied the coronal seismology technique to estimate the density and magnetic field variation at the location of open magnetic field regions. The nature of variation of magnetic field with radial distance was consistent with those predicted by the potential field source surface (PFSS) model. Similarly, \citet{2020ApJ...894...79W} has estimated the density gradients studying the height variation of the amplitude of the Alfv\'enic waves in the open magnetic field regions. The nature of variation of density matches well with those derived in \citet{2012ApJ...751..110B}. The observed slow variation of density leads \citet{2020ApJ...894...79W} to claim that there might exists an extended transition region due to the presence of spicules. \citet{2018ApJ...856..144M} have tested the accuracy of seismologic inversion to infer magnetic field strength in open magnetic field regions using MHD simulations and forward modelling of the propagating Alfv\'enic waves. They found that the accuracy of inversions does not depend on the fine structuring and choice of the spectroscopic lines. Recently, \citet{2020Sci...369..694Y} and \citet{2020ScChE..63.2357Y} observed prevailing transverse MHD waves with CoMP (which they identified as kink waves, which have an Alfv\'enic nature) and used these waves, in conjunction with estimates of electron number density, to map the plane-of-sky component of the global coronal magnetic field (found to be around 1 to 4 gauss, between 1.05 to 1.35 solar radii, where these results encompass both open and closed magnetic fields). Overall, these studies indicate that Alfv\'enic waves are a fundamental feature of the Sun's corona.

\section{Connection with solar wind acceleration}
\label{solar_wind}
Apart from heating the solar atmosphere and diagnosing the plasma conditions, Alfv\'enic waves play an important role in accelerating the solar wind. After the advent of space based observations, there has been a significant observational and theoretical advancement in this topic of research.

\subsection{Observational inputs}

Recent observations show that Alfv\'{e}nic waves are abundant at least at the bottom of the open-field coronal hole regions. 

The chromosphere of open-field regions is often filled with long spicules \citep{2012ApJ...759...18P}. High-resolution and high-cadence observations from Hinode/SOT have revealed the prevalence of swaying motions of spicules, which have been interpreted as Alfv\'{e}nic waves \citep{2007Sci...318.1574D}. These transverse waves have been found to reveal amplitudes of 10-25 km/s and periods of 100-500 seconds. The estimated energy flux is on the order of 100 W m$^{-2}$, which is comparable to that required to drive the solar wind (for more detail see \citet{2020SSRv..216..140V}). The authors claimed that the discovery of these waves supports the solar wind models based on dissipation of low-frequency Alfv\'{e}n waves. However, relatively higher frequency ($\sim$60 s) waves have also been detected in some spicules, which may support solar wind models based on high-frequency Alfv\'{e}n waves \citep{2009A&A...497..525H, 2012arXiv1207.6417Y, 2011ApJ...736L..24O}. 

Signatures of twisting and torsional motions have also been reported in spicules. From the chromospheric spectra taken by SST, \cite{2012ApJ...752L..12D} identified both swaying motions and torsional motions \citep[also see][]{2017NatSR...743147S}. The amplitude of torsional motion is often slightly higher than that of the transverse motion. A subsequent study using IRIS observations has largely confirmed the existence of these motions \citep{2014Sci...346D.315D}. Some of these torsional motions could be related to torsional Alfv\'en waves, which may be associated with significant heating of the chromospheric and coronal plasma.   

IRIS has performed direct imaging and simultaneous spectroscopic observations of the solar transition region for the first time. The high-resolution image sequences taken by IRIS have revealed the prevalence of intermittent small-scale transition region jets from network lanes in the quiet Sun and open-field coronal hole regions \citep{2014Sci...346A.315T,2016SoPh..291.1129N}. These network jets are the most prominent dynamic features in the transition region, and some of them are likely the heating signature of the chromospheric spicules \citep{2014ApJ...792L..15P,2015ApJ...799L...3R}, whereas others appear to be the smallest jetlets discovered from SDO/AIA observations \citep{2014ApJ...787..118R,2018ApJ...868L..27P}. \cite{2014Sci...346A.315T} found that these jets exhibit swaying motions similar to that found in chromospheric spicules. In addition, the simultaneously taken spectra of Si IV 1393.77 are found to be obviously broadened at the locations of these jets, demonstrating that the well-known nonthermal broadening of transition region lines \citep{1998ApJ...505..957C} is actually related to the network jets. As stated in their paper, unresolved transverse motions associated with kink waves and torsional Alfv\'en waves propagating along these jets are most likely responsible for the nonthermal broadening of the transition region spectral lines.  The wave amplitude was estimated to be $\sim$20 km/s, and the energy flux associated with the waves is 1-2 orders of magnitude higher than that required to drive the solar wind. 

\subsection{Theoretical modelling}
\label{section:tm_solwind}

The acceleration and heating of the solar wind are processes that are strongly coupled. \citet{Leer_1980JGR....85.4681L} realised this by noticing that the density of the wind at the critical point increases exponentially with the temperature at the base of the corona. Hence, the only way of generating a fast wind is to deposit the heat beyond the critical point. This fact naturally led to Alfv\'en waves becoming a favourite candidate for accelerating the solar wind (noting that the slow magnetosonic waves do not show sufficient energy flux in the solar  corona). This result has been further supported by the detection of Alfv\'enic fluctuations in-situ at distances 0.7-1AU by the Mariner 5 spacecraft by \citet{Belcher_1971JGR....76.3534B}, and Alfv\'enic waves permeating the base of the corona \citep{2007Sci...317.1192T,2015NatCo...6E7813M,2020Sci...369..694Y,2020ScChE..63.2357Y}. However, the latter observations do not detect sufficient energy flux in the Alfv\'enic fluctuations for coronal heating or solar wind acceleration, and the idea whether Alfv\'en waves and turbulence could account for solar wind acceleration and heating is still controversial \citep[e.g.,][]{2010ApJ...711.1044R}.

Indeed, Alfv\'en waves are notoriously difficult to dissipate at the base of the corona and the wave-action flux is expected to be constant with radial distance due to stratification and no dissipation (assuming radial expansion of the magnetic field in polar regions). This means that the non-linear factor $\delta B/B$ is expected to steadily increase with radial distance \citep{Jacques_1977ApJ...215..942J}. Hence, non-linear effects, which are a major candidate for Alfv\'en wave dissipation, become increasingly important at large distances  \citep[please see][for a review on solar wind models]{Hansteen_2012SSRv..172...89H}, while super-radial magnetic field divergence may counter this effect \citep[see the review,][] {2016SSRv..201...55A}.  In the nearly collisionless heliospheric solar wind plasma the eventual dissipation of turbulence and heating must occur on kinetic scales and is modeled typically with hybrid or Particle-In-Cell (PIC) or Vlasov's codes (for example, see the review \citet{2010LRSP....7....4O} and recently, \citep{2018ApJ...853...26F,2019SoPh..294..153R,10.3389/fspas.2016.00004,2019ApJ...879...53A,2019SoPh..294..114P,2019ApJ...887..208P}). 

Both, compressible and incompressible models for Alfv\'en wave dissipation have been successful in generating a corona and accelerating the solar wind. However, these models usually rely on empirical or WKB description of the waves, and the dissipation/heating processes are not modeled explicitly (i.e., the detailed physics is not included). Moreover, the perpendicular structuring of the plasma, intrinsic for the corona, which affects dramatically the MHD wave propagation, is usually neglected in these models. Regardless of the model the aims are the same: secure a turbulent cascade and/or the generation of small-scales down to kinetic dissipation scales in which the heating can take place in the nearly collisionless plasma. In the lower solar corona and in active regions collisions are more frequent, and MHD dissipation through viscosity and resistivity may be appropriate. One way to achieve this, particularly effective at the base of the corona and in closed loops, is the nonlinear mode conversion of Alfv\'en waves into longitudinal waves due to the centrifugal and ponderomotive forces (see \citet{2020SSRv..216..140V} for details). Since the area expansion factor directly controls the amplitude of these forces, numerical models have been successful at generating both the slow and fast solar wind only by changing this factor and was shown first using 2.5D MHD by \citet{1998JGR...10323677O}. Another popular mechanism is the Alfv\'en wave turbulence, which is discussed in the previous section, is based on the amount of wave-to-wave interaction in very low-beta plasma conditions. Alfv\'en wave turbulence can be achieved from linear processes as in the RMHD models \citep[e.g.][in which wave reflection is only achieved from the  change in Alfv\'en speed with height]{Velli_1993A&A...270..304V}, or nonlinear processes such as the parametric decay instability (PDI). 

An important constraint in models of incompressible Alfv\'en wave turbulence is the ratio of sunward to anti-sunward Alfv\'en wave energies. This ratio is observed to increase with distance \citep{Bavassano_2000JGR...10515959B}, from a value of $\approx0.3$ at 1.5 AU to $\approx 0.5$ at 4 AU, a fact that appears in opposition to Alfv\'enic turbulence due to the expected dynamic alignment effect \citep{Dobrowolny_1980PhRvL..45..144D}. Indeed, a daughter (reflected) wave is expected to decay faster than a parent wave, thereby leading to a dominant anti-sunward wave fraction and a corresponding decrease with radial distance of the sunward to anti-sunward ratio. 

The PDI is an instability of the Alfv\'en wave in compressible low beta plasmas \citep{Sagdeev_1969npt..book.....S,Goldstein_1978ApJ...219..700G} and applies to both circularly polarised or linearly polarised waves. Assuming that the original (or parent) wave is propagating rightward, due to interaction with a compressible wave the rightward propagating Alfv\'en wave decays into a rightward propagating acoustic wave and a leftward propagating Alfv\'en wave. This nonlinear reflection is also known as backscattering. \citet{Shoda_2016ApJ...820..123S} show that this kind of reflection due to compressible waves is substantial and allows to explain the observed increase in the ratio of sunward to anti-sunward Alfv\'enic fluctuations despite of the dynamic alignment effect. 

Another important dissipation mechanism for Alfv\'en waves in the solar wind is phase mixing (\citet{Heyvaerts_1983AA...117..220H}, see \citet{2020SSRv..216..140V} for details). This mechanism is efficient in the presence of large transverse density inhomogeneity such as that produced by the PDI. The effects of phase mixing on the dissipation of Alfv\'en waves due to background inhomogeneity in coronal holes were studied using 2.5D MHD and theoretical models in the past \citep[see the reviews][]{2005SSRv..120...67O,2016GMS...216..241O}. One of the first fully self consistent models of fast solar wind acceleration by Alfv\'en waves in a transversely inhomogeneous medium in 2.5D MHD was carried out by \citet{2012ApJ...749....8M}. 

\begin{figure}[!h]
\includegraphics[scale=0.4,angle=90]{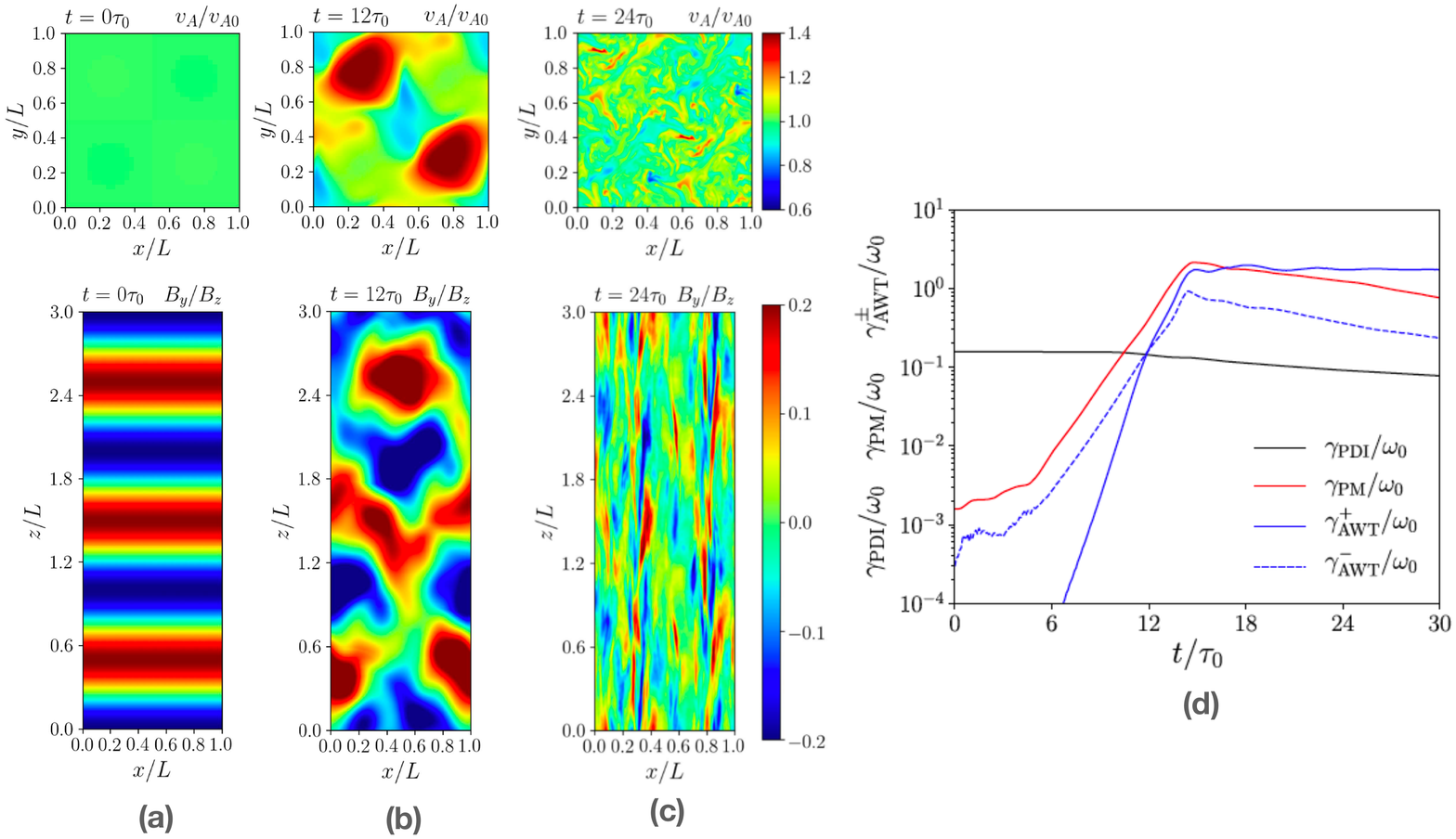}
\caption{Growth of parametric decay instability (PDI) with time. {\it (a)}: Circularly polarized Alfv\'en wave. Top panel shows a transverse slice and bottom panel shows a longitudinal slice. A random density fluctuation is imposed along z to trigger PDI. {\it (b)}: At 12$\tau_{0}$ (saturation phase), PDI has fully developed in simulations leading to large density imhomogeneities and phase mixing. {\it (c)}: At 24$\tau_{0 }$ (fully non-linear phase), Alfv\'en wave turbulence (AWT) is developed. {\it (d)}: Evolution of the growth rate of PDI, phase mixing, and Alfv\'en wave turbulence of forward and backward propagating (denoted by + and -) with time. Initially, PDI dominates the system. It leads to large transverse density inhomogeneities leading phase mixing dominate the system in the saturation phase. Finally, in the fully non-linear phase, AWT dominates the system. Adapted from \citep{Shoda_2018ApJ...853..190S}}
\label{fig-pdi}
\end{figure}

By conducting a first of its kind 3D MHD numerical simulation including compressible effects, \citet{Shoda_2018ApJ...859L..17S} show that the early nonlinear stage is dominated by phase mixing \citep[as the kind discussed by][]{2017NatSR...714820M}, thanks to the fast growth of the PDI (see Figure~\ref{fig-pdi} (b) and (d)). On the other hand, Alfv\'en wave turbulence dominates the fully nonlinear stage (see Figure~\ref{fig-pdi} (c) and (d)). It is therefore expected that regions with large density fluctuation (that may be attributed to the PDI), such as the coronal base and solar wind acceleration region \citep{2014ApJ...788..152R}, are likely dominated by phase mixing, while the other regions are dominated by Alfv\'en wave turbulence. The observed fast saturation of the non-thermal line broadening \citep{Hahn_2013ApJ...776...78H} may then be a consequence of the PDI-driven large density fluctuations at the coronal base.

Recently, quasi-periodic fast propagating (QFPs) waves associated with solar flares were discovered with SDO/AIA in EUV observations of coronal active regions \citep{2011ApJ...736L..13L}. These waves were identified as fast magnetosonic waves using 3D MHD modeling as well as 2.5D MHD model in a waveguide, and shown to carry significant energy flux for coronal heating \citep{2011ApJ...740L..33O,2014A&A...569A..12N,2018ApJ...860...54O}. On the other hand, using observations in the radio wavelengths \citet{2018ApJ...861...33K} detected signatures of QFP waves in the open coronal structure at about 1.7 R$_{sun}$, carrying the energy flux at least an order of magnitude lower than the local radiative losses. Since the fast magnetosonic waves can travel across the magnetic field, they can potentially transfer energy flux from flaring loops to open magnetic field regions. Since the initial discovery the QFPs were observed in many events. The observations of QFPs show that the waves damp within short distance (several wavelengths) from the flaring source in an active regions, that is likely due to the combination of magnetic field divergence and wave dissipation processes. At present, the statistics and occurrence rate of the QFPs in the corona is in initial stages of study \citep{2016AIPC.1720d0010L}, and it is not yet known whether the QFPs are produced at sufficient rate and amplitude for coronal heating.

The solar wind speed and the location of the heating are furthermore critically dependent on the photospheric perpendicular correlation length $\lambda_0$, which sets the length scale of the  dominant driver for the transverse waves. The solar wind speed generally increases with $\lambda_0$ \citep{Verdini_2010ApJ...708L.116V}, while the maximum temperature is almost constant with respect to it. As explained by \citet{Shoda_2018ApJ...853..190S}, this is because  shock heating and turbulence heating readjust. Indeed, a reduction in $\lambda_0$ leads to enhanced turbulence heating, which then increases coronal pressure and reduces compressibility, thereby reducing the efficiency of shock heating. Values of $\lambda_0 \lesssim 1$ Mm lead to dominant turbulence heating. However, compressibility still plays an essential role in this heating since it enhances the wave reflection via the PDI \citep{Shoda_2018ApJ...853..190S}. 

The value of $\lambda_0$ is determined by the properties of magnetoconvection, but its dominant spatial range is still a matter of debate. If it is determined by the photospheric kinetic energy peak then $\lambda_0\approx1$~Mm is expected \citep{Rempel_2014ApJ...789..132R}, case in which granulation (and thus the observed swaying motion of flux tubes) is the main driver. However, sub-arcsecond magnetic patches and vortex flows may be more energetically important, setting $\lambda_0\approx0.1$~Mm.

\section{Summary and Future directions}
\label{summary}
The progress made in recent years in our understanding of MHD waves as observed in open coronal structures is presented here. We particularly discuss the results pertaining to slow magnetoacoustic waves and Alfv\'enic waves alongside the role of latter in the acceleration of solar wind. The openness of coronal structures is determined based on the wave propagation lengths (with respect to the scale of the structure) but not on their magnetic connectivity. So, as such, the relevant research on propagating slow waves in active region loops is also discussed in this article.

It has been long known that the slow waves observed in active region fan loops originate in the lower atmosphere. Recent high-resolution, multi-instrument, multi-wavelength observations have provided a concrete evidence to show that they are actually generated via the leakage of photospheric oscillations into the outer layers \citep[e.g.,][]{2012ApJ...757..160J, 2015ApJ...812L..15K, 2016ApJ...830L..17Z, 2017ApJ...836...18C}. However, different schools of thought still exist on how these photospheric oscillations are generated. Some researchers find a connection with global $p$-modes while others demonstrate that they are locally generated through turbulent magnetoconvection. The possible location of the source region beneath the photosphere has also been discussed. Despite having similar properties, the existence of long period oscillations observed in polar regions is difficult to explain. Their origin has been studied in relation to transient events (spicules, reconnection jets, stochastic transients, etc.) occurring in the lower atmosphere \citep[e.g.,][]{2015ApJ...809L..17J, 2015ApJ...815L..16S, 2016ApJS..224...30Y, 2019Sci...366..890S}. While we are yet to find a solid evidence to pinpoint to a single source, the bigger challenge, in fact, is to clearly distinguish the PDs in terms of either slow waves or the recently discovered high-speed quasi-periodic upflows. Indeed, using simple models of slow waves and periodic flows, \cite{2015SoPh..290..399D} did not find any observational characteristics that would allow to distinguish between the two interpretations in isolation (i.e., without relying on a direct comparison between the two models) within the spatial, temporal, and spectral limits of the current instruments. On the other hand, \citet{2010ApJ...724L.194V} demonstrated that the observables which are usually considered as signatures of upflows could be easily explained in terms of the wave model.

Recent work has highlighted that the PDs observed in network plage regions might be more complex than a simple interpretation in terms of waves or flows. \citet{2005ApJ...624L..61D} and \citet{2006RSPTA.364..383D} already suggested a close relationship between PDs in the corona and (type II) spicules, based on the similarities in their properties. This was recently confirmed by \citet{2017ApJ...845L..18D}, based on combining SDO/AIA and IRIS observations with numerical simulations by \citet{2017Sci...356.1269M}. Their modelling results suggest that in plage region loops, PDs are the signatures of a complex series of events where the generation of spicular flows is linked to shock waves propagating into the corona and plasma heating derived from the dissipation of both waves and electric currents. PDs are not always small-amplitude perturbations of a pre-existing coronal structure but can actually represent the formation of new coronal strands. This complexity results in a mixture (and hence wide range) of properties, particularly the propagation speeds, as aspects of real flows, shock waves and current dissipation are all present, providing a natural explanation for the apparent contradictions in the interpretation of PDs in the literature. Although they are generally reported to only contain very modest energy budgets, PDs could play a significant role in the mass and energy flow in the solar corona through their close link with spicule-driven heating.

The rapid damping of slow waves has been extensively studied. It has been found that PDs in polar regions exhibit anomalous frequency-dependent damping (see Fig.{\,}\ref{kpfig3}) which is not compatible with traditionally considered dissipation mechanisms within the linear regime. It may seem that the damping lengths are underestimated especially for longer periods because of the instrument sensitivity or other detection limitations at large distances from the limb. However, the gradual decrease in damping length observed together with the fact that the damping lengths at longer periods are actually smaller than the possibly more reliable values at shorter periods makes these observational results convincing and therefore, requires an explanation. The temperature dependence of damping lengths in active region PDs also shows notable discrepancies with the theory (see Fig.{\,}\ref{kpfig4}). The results appear to suggest that thermal conduction is suppressed in hotter loops which is, although in agreement with previous observations, counter-intuitive. Further investigations are necessary to develop a proper understanding of this behaviour. The damping of slow waves due to the newly introduced thermal misbalance, is another interesting aspect to pursue further in the future.

A number of applications of slow waves that are useful to probe the plasma temperature, magnetic field strength, thermal conduction, polytropic index, and coronal heating function are discussed. While noting that there are still limitations to the individual techniques applied (e.g., projection effects), one may start exploring the scope of implementing the methods over large datasets.

Although indirect evidence of Alfv\'enic waves in the polar regions has been known since early 1970s, only after the advent of high resolution imaging and spectroscopic instruments such as AIA/SDO, EIS/Hinode, and CoMP, these waves could be resolved both spatially and temporally. Past observations and recent simulations have outlined the importance of studying the nonthermal line widths in establishing the prevalence of Alfv\'enic waves in the solar atmosphere and estimating the wave energy flux. The nonthermal line broadening in the corona is found to be correlated with the Doppler velocities, a feature that can be reproduced by the propagating kink/Alfv\'enic waves in the solar corona. However, in the transition region, IRIS observations have demonstrated the cause of nonthermal broadening is the network jets \citep{2014Sci...346A.315T}. Furthermore, the temperature dependence of the nonthermal broadening \citep{1998ApJ...505..957C} is still not well understood. Could it be related to different properties of Alfv\'{e}n waves at different temperatures? Furthermore the nature of variation of the nonthermal broadening with height is shown to be dependent on temperature. For example, while the nonthermal line widths of Fe X (red, 6347\AA) emission line increases with height, the nonthermal line widths of Fe XIV (green, 5303\AA) emission line decreases with height in the solar atmosphere at the locations of both open and closed magnetic field regions \citep{2002PASJ...54..793S, 2013SoPh..282..427K}. Our understanding is quite limited in explaining this nature of variation. Simultaneous multi-temperature and high-resolution spectroscopic observations of the chromosphere, transition region and corona, such as observations that will be made by the SPICE spectrometer on board Solar Orbiter, the proposed EUVST spectrograph, and the proposed VELC spectrograph onboard ADITYA-L1 may be required to answer these questions. 

There have been significant improvements in the numerical models and these have been somewhat successful in explaining a few observed properties of the solar corona. Most of these models assume a hot corona already present. Thus more efforts are needed to build a model that can self-consistently maintain the corona and reproduce the observed spectroscopic properties. Furthermore, the role of Alfv\'en(ic) waves in generating fine structures in the coronal plumes should be investigated in more details and compared with the observations. 

While MHD models could be appropriate for modeling waves and heating in the lower coronal plasma, where collisions are frequent, in the heliospheric plasma the collision are rare and the plasma and kinetic instabilities become important, as evident from in-situ observations at 1AU by Wind and other spacecraft, by Helios to 0.3AU and recently from Parker Solar Probe (PSP) in the inner heliosphere. Therefore, while MHD models could be applied to model the inertial range of turbulence and long wavelength MHD waves, kinetic models must be used to model important small-scale and kinetic processes, such as resonant dissipation of turbulence cascade, temperature anisotropies and related plasma instabilities (such as ion-cyclotron, and firehose), as well as likely Landau damping of compressional wave modes. Thus, the future direction of wave studies in open coronal structures must extend the modeling into kinetic regime in order to capture the physics of solar wind kinetic plasma processes, evident in spacecraft observations.

\begin{acknowledgements}
The authors thank the anonymous referees for their helpful suggestions. We wish to acknowledge ISSI-BJ for hosting the workshop at Beijing and for their generous Support for all the participants.
SKP is grateful to FWO Vlaanderen for a senior postdoctoral fellowship (No. 12ZF420N).
VP was supported by the Spanish Ministerio de Ciencia, Innovaci\'on y Universidades through project PGC2018-102108-B-I00 and FEDER funds. VP and TVD are supported by the European Research Council (ERC) under the European Union's Horizon 2020 research and innovation programme (grant agreement No 724326). 
J.A.M. acknowledges UK Science and Technology Facilities Council (STFC) support from grant ST/T000384/1. L.O. acknowledges support by
NASA grants NNX16AF78G, 80NSSC18K1131 and NASA Cooperative Agreement
NNG11PL10A to CUA. P.A. acknowledges funding from his STFC Ernest Rutherford Fellowship (No. ST/R004285/2). IDM acknowledges support from the UK Science and Technology Facilities Council (consolidated grants ST/N000609/1 and ST/S000402/1), the European Union Horizon 2020 research and innovation programme (grant agreement No. 647214) and the Research Council of Norway through its Centres of Excellence scheme, project number 262622. H.T. is supported by NSFC grants 11825301 and 11790304(11790300).
T.J.W. acknowledges support by NASA grants 80NSSC18K1131, 80NSSC18K0668, and NASA Cooperative Agreement NNG11PL10A to CUA.

\end{acknowledgements}

%
%

\bibliographystyle{aasjournal}      
\bibliography{references}   

\end{document}